\documentclass[useAMS,usenatbib]{mn2e}
\usepackage{graphicx}
\usepackage{float}
\usepackage{amsmath}
\usepackage{multirow}
\usepackage{bm}
\usepackage{lastpage} 
\usepackage{url} 

\newcommand{\mc}[3]{\multicolumn{#1}{#2}{#3}}
\newcommand{\msun}{~\text{M}_\odot}
\newcommand{\msunb}{\text{M}_\odot}
\newcommand{\xc}{\text{X}_\text{C}}
\newcommand{\xo}{\text{X}_\text{O}}
\newcommand{\tachar}[1]{
\setbox4=\hbox{\ }
\setbox3=\hbox{#1}
\hbox{#1}
\kern -\wd3 \kern -\wd4
\raise 0.45\ht3 \hbox{ \vrule width \wd3 height 0.5pt}
}
\newcommand{\tachard}[1]{
\setbox4=\hbox{\ }
\setbox3=\hbox{#1}
\hbox{#1}
\kern -\wd3 \kern -\wd4
\raise 0.35\ht3 \hbox{ \vrule width \wd3 height 0.5pt}
}
\newcommand{\rquad}{\tachar{I}}
\newcommand{\rquadot}{\tachard{$\dot{\text{I}}$}}
\newcommand{\rquaddot}{\tachard{\"{I}}}

\title[High resolution simulations of the head-on collision of WD]{High resolution simulations of the head-on collision of white dwarfs}
\author[D. Garc\'ia-Senz, R.M. Cabez\'on, A. Arcones, A. Rela\~no and F.K. Thielemann]{D. Garc\'ia-Senz$^{1,2}$\thanks{E-mail:domingo.garcia@upc.edu}, R.M. Cabez\'on$^3$, A. Arcones$^{4,5}$, A. Rela\~no $^1$ and F.K. Thielemann$^{3}$\\
$^1$Departament de F\'isica i Enginyeria Nuclear. Universitat Polit\`ecnica de Catalunya. c/Compte d'Urgell 187. 08036 Barcelona, Spain\\
$^2$Institut d'Estudis Espacials de Catalunya. c/Gran Capit\`a 2-4, 08034 Barcelona, Spain\\
$^3$Departement Physik. Universit\"at Basel. Klingelbergstrasse, 82. 4056 Basel, Switzerland\\
$^4$Institut f{\"u}r Kernphysik, Technische Universit{\"a}t Darmstadt, 64289 Darmstadt, Germany\\
$^5$GSI Helmholtzzentrum f\"ur Schwerionenforschung, Planckstr$\beta$e, 1. 64291 Darmstadt, Germany}

\begin{document}

\date{Some moment within 2013}

\pagerange{\pageref{firstpage}--\pageref{LastPage}} \pubyear{2013}

\maketitle

\label{firstpage}

\begin{abstract}
The direct impact of white dwarfs has been suggested as a plausible channel for type Ia supernovae. In spite of their (a priori) rareness, in highly populated globular clusters and in galactic centers, where the amount of white dwarfs is considerable, the rate of violent collisions between two of them might be non-negligible. Even more, there are indications that binary white dwarf systems orbited by a third stellar-mass body have an important chance to induce a clean head-on collision. Therefore, this scenario represents a source of contamination for the supernova light-curves sample that it is used as standard candles in cosmology, and it deserves further investigation. Some groups have conducted numerical simulations of this scenario, but their results show several differences. In this paper we address some of the possible sources of these differences, presenting the results of high resolution hydrodynamical simulations jointly with a detailed nuclear post-processing of the nuclear abundances, to check the viability of white dwarf collisions to produce significant amounts of $^{56}$Ni. To that purpose, we use a 2D-axial symmetric smoothed particle hydrodynamic code to obtain a resolution considerably higher than in previous studies. In this work, we also study how the initial mass and nuclear composition affect the results. The gravitational wave emission is also calculated, as this is a unique signature of this kind of events. All calculated models produce a significant amount of $^{56}$Ni, ranging from $0.1\msun$ to $1.1\msun$, compatible not only with normal-Branch type Ia supernova but also with the subluminous and super-Chandrasekhar subset. Nevertheless, the distribution mass-function of white dwarfs favors collisions among $0.6-0.7 \msun$ objects, leading to subluminous events.
\end{abstract}
\begin{keywords}
 Supernovae: general-- white dwarfs.
\end{keywords}

\section{Introduction}

There is a growing evidence supporting that the astrophysical scenario leading to a type Ia supernova (SNIa) explosion does not necessarily have to be unique. This assessment is probably true in light of the diversity shown by the light curves of SNIa as a function of the galaxy type; with the brightest events associated to galaxies with high star formation rate. The inferred amount of $^{56}$Ni from observations, points to the fact that two or even more different progenitors may exist \citep{howell2006,hicken2007,bianco2011}. Given the importance of SNIa in modern astrophysics, it is of great relevance to discern between the various potential channels that produce them.

Accretion onto massive white dwarfs (WD) composed of carbon and oxygen is one of the most favored mechanisms to account for type Ia supernova explosions (see \citealt{hillebrandt2000}, and references therein). Nevertheless, there is more than one way to trigger the thermonuclear explosion of a white dwarf. This gives rise to different scenarios, that are divided into two main categories, depending on how many white dwarfs are involved: the single WD scenario \citep{whelan1973,nomoto1982} and the double-degenerate progenitor \citep{iben1984}. In the first one, a WD accretes material from a non-degenerate companion until the Chandrasekhar mass is  approached. Then, the binding energy of the WD becomes marginally negative~while the temperature at its center gets high enough to allow the off-center carbon ignition under degenerate conditions. A nuclear flame settles spreading all along the WD, and in a couple of seconds, practically the entire star is burnt, leading to its disruption (see \citealt{nomoto1984, woosley1986} for early studies and \citealt{roepke2012} for a more recent approach). In the second scenario two WDs collide \citep{pakmor2010}. This collision can occur either by the loss of angular momentum by gravitational waves emission and unstable mass accretion in the coalescence scenario \citep{dan2011}, or by the more violent head-on encounter \citep{katz2012}. This last case  has been studied by \cite{benz1989, rosswog2009.2, raskin2009, raskin2010, hawley2012}~being also the matter of study in this work\footnote{See also the recent manuscript by \cite{kusnhir2013} simulating the head-on scenario with an Eulerian axisymmetric code.}

A rough estimation of the direct collision rate for WDs in a typical globular cluster and in the galactic center was done by \cite{benz1989}. For a typical globular cluster they inferred a total of $\simeq 120$~white dwarf collisions in the galaxy within the Hubble time. The number of collisions in the galactic center was estimated an order of magnitude higher, $\simeq 1,870$. Taking both contributions and, assuming that all the collisions end in an explosion, gives an upper limit for the  rate of this channel: $\simeq 2\cdot 10^{-7}$~events/yr, considerably less than the observed  SNIa rate $\simeq 2\cdot 10^{-3}$~yr$^{-1}$ \citep{cappellaro1999}. Nevertheless, these rates might have been underestimated under the light of the work of \cite{katz2012}. They showed that binary WD systems that are orbited by a distant perturber have a few percent chance of a clean head-on collision within 5 Gyr, by calculating approximately ten thousand 3-body integrations for a wide range of initial conditions, and different kinds of perturbers.

The collisional channel merits to be investigated at least by two reasons. First, it will be useful to clean the template sample of standard SNIa from outliers in order to reduce the dispersion. The second reason is that this channel provides a firm mechanism to get high amounts ($> 0.7\msun$) of $^{56}$Ni, hard to obtain in other scenarios. The much lower rate observed for these super-Chandrasekhar mass explosions could be more easily conciliated to the expected rate of the collisional channel.

The numerical studies of the head-on scenario were pioneered by \cite{benz1989}, and have been revisited by \cite{rosswog2009.2}, \cite{raskin2009, raskin2010} and \cite{hawley2012}. In those works, the supernova is triggered by the direct encounter of a pair of WDs with masses in the range $0.4\msun \leq M_{WD}\leq 1.1\msun$. The results of these studies vary from low $^{56}$Ni production ($\simeq 0.01 \msun$) to large yields ($\simeq 1 \msun$). The most complete study to date was that carried out by \cite{raskin2010}, using a 3D smoothed particle hydrodynamics (SPH) code, where they explored different WD masses, mass ratios and impact parameters. The main conclusion of the study was that, for a reasonable choice of parameters, it was not difficult to obtain $0.3-0.8\msun$ of $^{56}$Ni.

In this work we use a 2D axial-symmetric SPH code to simulate the head-on collision of two WDs along the symmetry axis (i.e. with impact parameter $b=0$). The motivation is to rule-out resolution problems as the cause for the production (or absence) of high $^{56}$Ni yields, which might partially explain the different results obtained among previous works. We also post-processed the nuclear abundances obtained from the numerical simulation, using the extended BasNet reaction network \citep{thielemann2011}. In this way, we want to validate the results obtained with the simple alpha-network, that is typically used in hydrodynamical calculations. A comparison between models 2D7755 and 3D7755 in Table 1 indicates that the maximum resolution achieved by {\sl AxisSPH} using $N=90,000$ particles ($\simeq 40$~km for model 2D7755 at t=0 s) is much better than that of current 3D calculations using several hundred thousand particles ($\simeq 150$~km for model 3D7755 at t=0 s). This allows to invest more computational effort in the combustion process itself by taking tiny time-steps, often smaller than $1~\mu s$. In the direct collision scenario, the nuclear combustion takes place, for the most part, in the quasi-statistical equilibrium regime (QSE)~and the complete statistical equilibrium of the species can not be assumed (as in the Chandrasekhar-mass models of SNIa). Instead, individual reactions have to be tracked in detail using very small time steps. Thus, in practice, it is necessary to compute a large number of models to reasonably describe the nuclear combustion. The main aim of this work is to determine the main features of these explosions and the nucleosynthetic yields performing numerical simulations of the head-on collisions of white dwarfs with high {\sl spatial} and {\sl temporal} resolution. The axisymmetric hypothesis helps to make the calculation feasible.

This paper also brings information concerning other parameters which were not addressed in previous works. For example, we explored the sensitivity of the ensuing nucleosynthetic yields with respect to the initial carbon content of the white dwarfs. The gravitational radiation resulting from this kind of stellar encounters is also given, as it is an important distinct signature to identify these progenitors in the future.

The paper is organized as follows. In Section~\ref{setting}, we describe the numerical setting that we used to perform the simulations and the nuclear network used for the post-processing (Section~\ref{network}). In Section~\ref{results}, we present the results of the calculations, that include the description of the overall evolution of the nominal case of $0.7\msun +0.7\msun$ (Sections~\ref{0707} and \ref{2Dhighres}), the impact of the WDs masses (Section~\ref{mass}), the dependence of the results on the initial carbon abundance (Section \ref{carbon}), the comparison of the results with the post-processed $^{56}$Ni yields (Section~\ref{postprocess}) and the calculation of the resulting gravitational wave emission for the $0.7\msun +0.7\msun$ and $1.06\msun + 0.81\msun$ scenarios (Section~\ref{gwave}). Finally, Section~\ref{conclusions} is devoted to the conclusions and final remarks.

\section{Method}
\subsection{Numerical setting}
\label{setting}
In terms of computational power, it is too costly to improve the resolution in 3D simulations beyond previous works and up to a level that really makes a difference. Thus, we decided to use an axisymmetric SPH code and study the $b=0$ collision with very high resolution. All calculations were carried out with the code {\sl AxisSPH}, developed and detailed in \cite{garciasenz2009} and \cite{relano2012}.  Given the importance of the hydrodynamic method used to carry out the simulations we give a summary of the mathematical formulation of the numerical scheme as well as of their most relevant features in the Appendix~\ref{appAxisSPH}.

We also simulated a 3D low-resolution collision of two twin $0.7\msun$ white dwarfs to compare the results with the 2D high-resolution calculation and corroborate previous convergence studies. To that purpose, we used our 3D-SPH code \citep{garciasenz2005,bravo2008} conveniently adapted to handle the present scenario.

Both codes include an equation of state (EOS) consisting of radiation plus a mixture of ions, treated as an ideal gas with Coulomb corrections including plasma polarization effects, and relativistic electrons with any degree of degeneracy. The energy release due to nuclear combustion is taken into account using a 14-nuclei alpha-network (from $^4$He to $^{60}$Zn) that also includes the binary reactions $^{12}$C~$+^{12}$C, $^{12}$C~$+^{16}$O and $^{16}$O~$+^{16}$O \citep{cabezon2004}. One key ingredient of this network is that the evolution of nuclear species at each time-step is implicitly coupled with the  
temperature evolution. The value of both, the molar fractions of the species and temperature are simultaneously found at step $n+1$~as a function not only of their known values at the step $n$~but also on their final, unknown values, at step $n+1$.  Starting from the values $y^n$~and $T^n$~of the molar fractions and temperature this is done by solving the coupled set of equations $dy/dt= F(y^{n+1}, T^{n+1});~dT/dt= G(y^{n+1}, T^{n+1})$~with the Newton-Raphson technique, where $F$~describe the evolution of the fourteen nuclei belonging to the $\alpha-$chain and $G$~is the energy equation (or, equivalently the temperature equation). This feature allows to handle in a very robust way the combustion stages of the mixture, from the nuclear-statistical equilibrium (NSE) regime to normal combustion, including the quasi-statistical equilibrium (QSE) and final freeze-out of the species. A network of similar size was used by \cite{rosswog2009.2} but without the implicit coupling to the temperature evolution. \cite{raskin2010} used a hybrid hydrostatic/self-heating network to compute the nuclear combustion.

\begin{figure}
\includegraphics[angle=-90,width=\columnwidth]{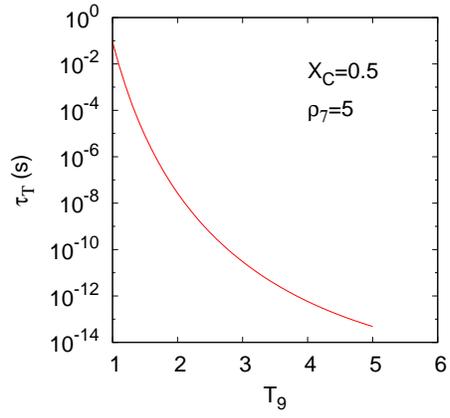}
\caption{Characteristic thermal timescale $\tau_{T}$ for $^{12}$C combustion via the $^{12}$C$+^{12}$C reaction at a density of $\rho=5\cdot 10^7$~g/cm$^3$ and with carbon mass fraction $\xc=0.5$.}
\label{fig01}
\end{figure}

Actually, the correct handling of nuclear reactions is the most challenging piece of this kind of calculations. To understand why, Fig.\ref{fig01} depicts the characteristic time $\tau_T$ of temperature increase due to the $^{12}$C~$+^{12}$C reaction at a constant density $\rho=5\cdot 10^7$~g/cm$^3$. This time is here defined as $\tau_T=\left(\frac{c_v~T}{\dot S_{nuc}}\right)\eta$ (i.e. the time step leading to an increase of temperature by the relative fraction $\eta$) being $\dot S_{nuc}$ the instantaneous nuclear energy generation rate, $\eta=10^{-2}$, and $c_v$ the specific heat at each temperature, computed using the EOS. As we can see, when temperature exceeds two billion degrees the time step falls down to $0.01~\mu s$ which makes it difficult to handle the hydrodynamics jointly with the nuclear burning in multidimensional calculations. This means that even using an $\alpha$-network, it is necessary to adopt a clever approach to handle the combustion. A good starting point is to consider the implicit coupling between nuclear reactions and temperature via the energy equation \citep{mueller1986,cabezon2004}, because it stabilizes the combustion, especially when photodisintegrations take over. Nevertheless, the implicit thermal coupling is not a sufficient condition to make the calculation feasible in practice, owing to the very small time steps. Thus, we have made use of two additional hypotheses. First, when the current hydrodynamical time step becomes larger than the instantaneous thermal time $\tau_T$ we make use of an operator splitting technique. In this situation the combustion algorithm decouples the nuclear network from the dynamics until the current hydrodynamical time step is recovered. This procedure allows to compute the energy release using effective time steps close to $10^{-10}$~s, small enough to describe the nuclear combustion of most of the models shown in Table~\ref{table1}. Our second approach concerns massive white dwarfs (i.e. only in our model 2D10855), where the central density is high and $\tau_T~(\propto \rho^{-2}$) becomes too small even for the splitting technique. Only for this run, we allow the material to jump to complete NSE if $\rho\ge 5\cdot 10^7$~g/cm$^3$ and $T\ge 3.5\cdot 10^9$~K. Our NSE routine has the same 14 nuclei as the $\alpha$-network, with exactly the same nuclear parameters (such as partition functions and binding energies) so that they are totally compatible. Once the transition to NSE has been completed, the further evolution of nuclear abundances is again calculated with the implicit $\alpha-$network, so that the freeze-out of the species is computed in a realistic way. Even with these simplifications the number of iterations needed to simulate an explosion is rather large, ranging from $\simeq 2.5\cdot 10^4$ for case 2D6637 to $\simeq 4\cdot 10^5$ for case 2D10855 (see Table~\ref{table1} for further details on the names of the runs).
            
In order to simulate the collision, we built 2D and 3D initial models of the WDs, being properly relaxed to ensure good mechanical equilibrium before the simulation starts. We place both stars with an initial separation $d_{12}^0$~(see Table 1) and give them an initial velocity corresponding to point-mass free-fall from infinity, up to that distance. In many of the the 2D simulations we used $N_{2D}=88,560$ particles. It is worth to note that to achieve the same resolution in 3D, roughly $N_{2D}^{3/2}\simeq 2.6\cdot 10^{7}$ particles would be needed. In order to have a similar resolution as in previous works, we used 200,000 particles for the 3D simulation, same as \cite{raskin2010}. Table~\ref{table1} shows a summary of the main features of the initial models used in this work.

\subsection{Extended nuclear network and post-processing}
\label{network}
In general, the alpha network used in the simulations provides only a rough estimation of the energy generation and to the amount of the most abundant isotopes, \cite{timmes2000}. However, a more accurate calculation of the nucleosynthesis requires a  more extended reaction network. In a post-processing step we have compared the results of the alpha network in the simulations with those obtained from the BasNet network (see e.g., \citealp{thielemann2011}), that includes in the present application more than 1,500 nuclei and all relevant reactions. We use REACLIB compilation, containing theoretical statistical-model rates by \cite{rauscher2000} (NON-SMOKER) and experimental rates by \cite{angulo1999} plus further updates (https://groups.nscl.msu.edu/jina/reaclib/db/). Where available, experimental beta-decay rates are used from \cite{nudat22009} supplemented by theoretical ones \citep{moeller2003}.  The theoretical weak interactions rates are taken from \cite{langanke2001}.

The Lagrangian nature of SPH allows all particles used in the calculation to behave as tracers from the nucleosynthetic point of view. Nevertheless, instead of running the extended network over the whole sample of particles, we select a smaller but representative set of particles for the post-processing nucleosynthesis calculations, utilizing the density and temperature evolution from the hydrodynamic (SPH) simulation. Their initial composition is thus $X(^{12}C)=X(^{16}O)=0.5$. For selecting these representative tracer particles the full set of particles is ordered with respect to the maximum temperatures encountered during their evolution, which functions as a criterion for the expected nucleosynthesis features. We used 21 bins, from T$_9=2$ up to 6 in intervals of $\Delta$T$_9=0.2$, with the exception of the last bin which ranges T$_9=6-7$. From each temperature bin we randomly select 50 particles, obtaining a total of 1,050 trajectories. We tested whether this procedure assures convergence in the abundances obtained by performing the nucleosynthesis postprocessing also for all tracers in two temperature bins, one responsible for the major production of intermediate-mass elements, and one for the dominant production of Fe-group elements. The result is that we find at most a 5\% deviation in abundances of relevance in this zone. This error can go up to 30\% for elements with a negligible abundance in the relevant mass zones. Thus, when integrating over all ejecta, a maximum error of 5\% for the composition is assured, in agreement with \cite{seitenzahl2010}, which is suitable for the scope of this paper.

\section[]{Results}
\label{results}

\subsection{The baseline case: collision of two $0.7\msun$ white dwarfs (Models 2D7755 and 3D7755)} 
\label{0707}
The baseline case simulates the head-on collision of two $0.7\msun$ WDs with an initial composition of $^{12}$C and $^{16}$O at equal parts. The numerical setting is made as described at the end of Section~\ref{setting}.

The sequence of events is summarized in the series of snapshots depicted in Fig.~\ref{fig02}, and its continuation in Fig.~\ref{fig03}, where we plot all particles from the 2D7755 run. The color represents, from left to right, density ($10^7$~g/cm$^3$), temperature ($10^9$~K), mass fraction of $^4$He $+^{12}$C $+^{16}$O, mass fraction of intermediate-mass elements (IME, from $^{20}$Ne to $^{40}$Ca) and mass fraction of Fe-group elements (from $^{44}$Ti to $^{60}$Zn). It has to be noted that the paucity of particles close to the z-axis in Figs.~\ref{fig02} and \ref{fig03}~is a natural consequence of the axisymmetric hypothesis. A point in the 2D-cartesian plane is a ring in the 3D space. At the same distance to the center of each star a volume element at the equator is more massive than the same element near the pole. This produces a dilution of the mass points near the symmetry axis when particles with the same mass are used to describe the hydrodynamics.     

The simulation starts before what is represented on Figs.~\ref{fig02} and \ref{fig03}. The main parameters of the configuration at t=0 s are given in Table~\ref{table1}. When the stars get closer their shape is deformed by tidal forces. Nevertheless, this deformation is not strong because of the short time available before contact. At $t=4.66$~s, both stars reach contact and the collision starts. At that time, the relative velocity of the stars is $v\simeq 4,800$~km/s, higher than the speed of sound, even at their center ($c_s\simeq 4,000$~km/s). A shock wave is formed at the contact region with increasing density and temperature. In the first row of Fig.~\ref{fig02} ($t=5.63$~s), the shock moves slowly outwards from the contact interface, until it stalls due to the ram kinetic pressure exerted by the fresh infalling material.  The stagnation of the accretion shock has been also described not only in other papers dealing with the collision scenario, \citep{raskin2010, rosswog2009.2}~but also in the abundant literature devoted to Type II supernova explosions  (see for example \citealt{woosley1986}, and references therein). The matter that has gone through the shock has increased its density and temperature, leading to combustion. As soon temperature reaches $10^9$~K and density $\simeq 10^7$~g/cm$^3$, $^{12}$C $+^{12}$C reaction settles around the collision plane, producing alpha particles that react with $^{12}$C and $^{16}$O, climbing up through the alpha-network, producing intermediate mass elements (second row of Fig.~\ref{fig02}, $t=6.19$~s). As it can be seen, the combustion is limited to a small region along the contact interface, because those particles, with high enough temperature, still have a relatively low density. At the typical peak densities reached during the collision the conductive and radiative heat transport processes are inefficient to evacuate the thermal energy accumulated in the detonated zone before disruption, \citep{raskin2010}. In consequence the energy liberated by nuclear reactions cannot leave the pocket of ashes that is left behind the shock, because it is surrounded by higher density regions, that induce a steep positive gradient of pressure in all directions. The combustion region remains then confined during a time $t\simeq 1$~s, large enough to allow the synthesis of approximately $0.06\msun$ of IME. 

At approximately $t=6.30$~s (third row of Fig.~\ref{fig02}), the temperature and density at the edge of the shock front are high enough to initiate carbon burning. In that region, energy can be freely released, giving rise to a detonation front that propagates all along the stalled shock front.  This detonation is so fast that usually emerges almost simultaneously from single particles and induces a rapid spontaneous burning, especially for those particles close to/at the stalled shock\footnote{Is still unclear whether the resolution is enough or not to resolve the initiation of the detonation (even more when dealing with the axisymmetric hypothesis), but the convergence tests reported in this work suggest that the main observables deduced from the simulations are robust. On one hand, the detonation wave that propagates into the low-density unburnt material (region out of the stalled shock), produces a big amount of IME ($\simeq 0.8\msun)$}. On the other hand, the wave propagating through the already hot burnt material rises the density and temperature of that material even more, producing Fe-group elements. Because the system is not absolutely symmetric (the stars are not initially mirrored, but displaced) the detonation may start at slight different times in both hemispheres. To quantify the impact of the initial degree of symmetry in the final outcome, a model was run starting from a totally symmetric initial configuration. The evolution of this model was similar to that of model 2D7755 but some differences are worth to comment. The symmetric case gives $0.33\msun$~of $^{56}$Ni (2D7755 produced $0.36\msun$), and a bit more  $^{28}$Si, $0.42\msun$, ($0.408\msun$ from model 2D7755) so that the energetics of the explosion of the mirrored and displaced cases remained similar. The absolute temperature peak achieved during the collision was ia little higher for the symmetric model ($\simeq 8\%$) but the density at the temperature peak was lower ($\simeq 6\%$). A slightly higher combustion temperature and reduced density may be the cause of the nucleosynthetic discrepancies between both models, as photodisintegrations are most favored by high temperatures and lower densities.  We should keep in mind, however, that starting from a totally ordered set of points is not always advisable when working with a particle code. An excessive symmetry may produce numerical artifacts when a particle meets with itself. A little amount of asymmetry is often beneficial in SPH,  being  this the reason behind the initial setting of models shown in Table  \ref{table1}.      

At $t=6.39$~s (first row of Fig.~\ref{fig03}) both detonation fronts have overcome each other, shocking again the burnt material of the contact region, and producing even more Fe-group elements, up to an amount of $0.38\msun$. From those, $0.36\msun$ are $^{56}$Ni. In the second row of Fig.~\ref{fig03} ($t=6.80$~s) the release of nuclear energy is large enough to outpower the gravitational potential and the kinetic pressure of the infalling material, leading to a rapid expansion of the star. 

At $t=12.95$~s (last row of Fig.~\ref{fig03}) the system is disrupted, and the decline of temperature and density freezes the nuclear abundances in a sort of distorted onion-layer structure, with lighter elements in the outer shells and heavier elements in the internal ones. Table~\ref{table2} shows the total energy of the gas: $1.62\cdot 10^{51}$~erg after freezing, similar to the $1.66\cdot 10^{51}$~erg obtained by \cite{kusnhir2013}~for the same scenario, and a little higher than $1.3\cdot 10^{51}$~erg of  model W7 by \cite{nomoto1984}. 

 One advantage of using a lagrangian method to compute the hydrodynamics is that the evolution of individual mass elements can be easily tracked. In Fig.~\ref{fig04} we represent the evolution in the plane $\rho-T$ of four selected particles of the upper white dwarf depicted in Figs.~\ref{fig02} and \ref{fig03}. Roughly speaking, the particle sample first undergoes a phase of adiabatic compression, until the temperature exceeds a billion degrees. Afterwards, nuclear release takes over and temperature rapidly climbs to $3-5.5$~GK, the precise value depending on the initial location of the particle. Nevertheless, the nuclear ashes do not expand, and remain stagnated and confined by the kinetic pressure of the infalling material. Once the released nuclear energy is large enough to overcome the kinetic pressure of the material, it gradually settles in the long adiabatic expansion line crossing Fig.~\ref{fig04}, until their total dilution. Particles 1 and 2, have a richer behavior than that of particles 3 and 4 in the upper part of their trajectories, because they describe additional small loops during the combustion phase (see Fig.~\ref{fig04}, lower panel). These loops point to the existence of trains of shock waves moving and reflecting within the detonated volume, modifying the thermodynamic trajectory of the particles. The analysis of the trajectories of similar mass elements in model 2D7755Res, calculated with higher resolution, supports this conclusion. 

We also conducted a 3D calculation of the same scenario using the same input physics and initial setting, in order to compare with the results of the 2D code and perform a convergence test. We used 200,000 particles for run 3D7755, so resolution is roughly a factor 5 lower than in run 2D7755.

The evolution in 3D is very similar to that calculated in 2D using {\sl AxisSPH}. We show the evolution of internal, kinetic and gravitational energies and the nuclear species in Fig.~\ref{fig05}, in comparison with those of run 2D7755. Despite the different code conception and resolution, the outcome of the simulations in 2D and 3D is strikingly similar. Minor differences can be seen in the evolution of energies and abundances. The production of IME is a bit delayed in the 2D calculation, which can be due to the lower value of the smoothing length in this run (corresponding to the higher resolution). This delays the contact between both stars in about $\delta t\simeq 4h/v_{rel}$, being $v_{rel}$ the relative velocity at the impact moment. Run 3D7755 synthesizes a bit more elements of the Fe-group, while the total amount of IME is practically equal in both calculations. The energetics of both simulations are consistent with the nucleosynthesis trend: the 3D calculation has a slightly higher final kinetic energy than the 2D run. The total amount of synthesized $^{56}$Ni is $0.39\msun$ in 3D and $0.36\msun$ in 2D (see Table~\ref{table3}).

Therefore, the comparison between the 2D and 3D calculations supports the conclusion of \cite{raskin2010} that $\simeq 2\cdot 10^5$ particles in 3D would be enough to capture the main features of the collision.

It is important to take into account that these limited nuclear-networks that are used coupled with hydrocodes provide a reasonable energy generation rate, but the final abundances should not be taken as being precise values. In order to obtain more realistic yields, we would need either a more extended network, that is prohibitively expensive when coupled with an hydrodynamical calculation, or to post-process the final result using the thermodynamical trajectories of the particles. Hence, in order to gain insight in the final outcome of the head-on collision scenario we post-processed the evolution with a larger nuclear network and compared the results in Section~\ref{postprocess}. 

\subsection{Convergence of the results: model 2D7755Res}
\label{2Dhighres}

Model 2D7755Res simulates the same scenario than model 2D7755 but using twice the number of particles (177,120). In this way, we increase the spatial resolution and, more importantly, the mass resolution by a factor 2. Then, we can compare the results between both calculations and establish a convergence of the outcome with resolution within the 2D approximation. The evolution of the collision is depicted in Figs.~\ref{fig02res} and \ref{fig03res}, following the same color profiles than in Figs.~\ref{fig02} and \ref{fig03}, and showing the snapshots at similar times. Comparing the results, it is clear that the general evolution is very similar, with the exception that the slight lack of symmetry in the early time of the detonations evolution, present in model 2D7755, is suppressed in model 2D5577Res. This is probably due to the better mass resolution, which allows us to follow more accurately the triggering of the detonations. The results of the simulation can be found in Tables~\ref{table2} and \ref{table3}. Model 2D5577Res generates a slightly higher amount of $^{56}$Ni ($0.364\msun$ in front $0.362\msun$ of model 2D7755), and a very similar amount of IME ($0.815\msun$ in front $0.808\msun$ of model 2D7755). As a consequence, the energetics are also very similar, and the final kinetic energy at infinity is almost the same ($1.63\cdot 10^{51}$~erg compared with $1.62\cdot 10^{51}$~erg of model 2D7755).

Under the light of these results we can conclude that the simulations using 88,560 particles, do a reasonable good work in estimating the final nucleosynthetic and energetic production.

\subsection{The impact of WD mass (Models 2D6655 and 2D10855)}
\label{mass}

According to Table \ref{table2}, changing the mass of the white dwarfs has a sizeable impact in the strength of the explosion. For a pair of canonical white dwarfs with mass $0.6\msun$ the final kinetic energy is $1.33\cdot 10^{51}$~erg, fully compatible with the kinetic energy of normal SNIa events. The energy of the explosion amounts to $1.62\cdot 10^{51}$~erg for twin $0.7\msun$ collisions, and to $2.06\cdot 10^{51}$~erg for the $0.81\msun +1.06\msun$ impact. This last case would give rise to a considerably bright event, observationally cataloged as a super Chandrasekhar-mass explosion. The amount of radioactive nickel ejected in the explosion also follows a similar trend going from low, $0.25\msun$, for model 2D6655 in Table~\ref{table3}, to high, $1.02\msun$, for model 2D10855. Therefore, near head-on collisions of canonical ($\simeq 0.6\msun$) white dwarfs would give rise to a subluminous SNIa explosion, in agreement with other studies \citep{raskin2010,rosswog2009.2}.

Model 2D10855 is the only super Chandrasekhar-mass model considered in this paper. Admittedly, even supposing a high rate of head-on impacts they would, for the most part, be the product of canonical $0.6-0.7\msun$ white dwarfs. The direct encounter of two white dwarfs with masses as high as $0.81\msun$~and $1.06\msun$~would be rare, with a low probability to be realized in nature. Still it is an interesting limiting case worth to explore. In Fig.~\ref{fig06} there are represented different time slices, similar to those of model 2D7755 presented in Section~\ref{0707}.

In this scenario, first contact happens at higher speeds than in case 2D7755. Nevertheless, due to its higher densities, the collision occurs subsonically for the most massive star, while it is super-sonic for the less massive one. This immediately triggers an asymmetric detonation shock that expands through the upper star (Fig.~\ref{fig06}, first and second rows). IME and Fe-peak elements are created in the less massive star while its companion remains without much alteration apart from a pressure wave that slowly compresses and heats the material, but not enough to switch on the nuclear reactions yet. In the third row of Fig.~\ref{fig06}, two new detonation fronts can be seen in the most massive WD. The first one is formed when the pressure wave crosses the center of the massive WD, while the second one happens close to the oblique contact layer between both stars (indicated by a red arrow). Both detonations shock the unburnt material twice, rising the temperature and the density enough to trigger a very fast carbon burning, and leading to a creation of an enormous amount of Fe-peak elements (first and second rows of Fig.~\ref{fig07}). These late detonations cross over the whole system, catching up with the initial fronts and releasing energy enough to disrupt the stars (third row of Fig.~\ref{fig07}). Interestingly, layers with different compositions start to mix in the last stages, due to the shape of the contact layer between both stars and the subsequent interaction between layers moving at different velocities.

\subsection{Dependence on the initial carbon abundance (Models 2D6637 and 2D6673)}  
\label{carbon}

The nuclear composition of the C+O white dwarfs participating in the explosion is often taken uniform with mass fractions $\xc =\xo =0.5$. Nevertheless, the precise composition profile along the star could also be a function of the mass of the white dwarf, and of the not yet well-known $^{12}$C$(\alpha, \gamma)^{16}$O reaction rate. In order to explore the sensitivity of the results with the initial carbon abundance, we have simulated two additional explosions varying this parameter: cases labeled 2D6673 and 2D6637 in Tables~\ref{table1}, \ref{table2} and \ref{table3}. The results of the simulations are summarized in Table~\ref{table2} and Fig.~\ref{fig08}. The most remarkable conclusion is that there is not a uniform trend in the results as the initial abundance of carbon changes, a clear signature that we are facing a highly non-linear phenomenon. For example, the lowest final kinetic energy corresponds to model 2D6637 ($\xc=0.3$) but the kinetic energies of cases  $\xc=0.7$~ and $\xc=0.5$~are virtually the same. Figure~\ref{fig08} represents the evolution of four nuclear groups: $\alpha$, C+O, IME and Fe, for the three choices of the initial carbon abundance. As we can see, the larger the initial carbon abundance is, the earlier the synthesis of the different groups begin. There is a considerable delay in the time at which the Fe nuclei begin to be synthesized, of almost one second, between models 2D6673 and 2D6637. Thus, for the $\xc=0.3$ model combustion takes place, on average, at higher densities than for model with $\xc=0.7$, leading to a stronger detonation but smaller abundances of IME. Regarding the $^{56}$Ni yields, these two effects (strength of the detonation and amount of IME produced) work in opposite directions, partially compensating mutually. The stronger detonation in the case of the system with lower amount of $^{12}$C, should lead to a high amount of $^{56}$Ni, but the yields are partially suppressed by the lower amount of seeds ($^{12}$C) for creating IME. This might explain why an intermediate case (like 2D6655) is more efficient producing $^{56}$Ni than the other two more extreme cases, where one of the two mechanisms dominates.

The uncertainties in the initial carbon abundance lead to a new scatter in explosion energies and nickel yields, to be added to those due to the initial mass of the white dwarfs and unknown impact parameter. Note that, for a given WD mass, the scatter in the yields of $^{56}$Ni caused by the initial $^{12}$C content is, in many cases, of similar size as the dispersion shown in Table \ref{table4} summarizing the calculations reported by other groups.  

\subsection{Comparison with an extended network}  
\label{postprocess}
We compare the results of the alpha network and the post-processed nucleosynthesis for model 2D7755 (baseline case, Sec.~\ref{0707}). In the post-processing calculation the amount of nickel ejected is 0.40~$\msun$, compared to $0.36~\msun$ from the alpha network in the simulation. This difference mainly comes from the opening of new channels which pump matter from IME to iron group, due to the combined action of $(\alpha,p)(p,\gamma)$ reactions which are typically faster than the $(\alpha,\gamma)$ reactions at very high temperature. A similar conclusion was reached by \cite{timmes2000} in their comparative study of nuclear networks. We have also checked that the large and the reduced networks provide similar amounts of released nuclear energy. For the total, time-integrated, nuclear energy released up to the freezing of the abundances the relative difference between both calculations is $\simeq 3\% $. Although electron captures are also included in the post-processing network, they are negligible at the typical densities reached during the collision. 

Fig.~\ref{fig09} shows a comparison of the abundance evolution for one particle calculated with the full postprocessing network (dashed lines) and the reduced alpha network employed in the simulations (solid lines). This specific particle experiences temperature conditions which are at the borderline of explosive O-burning and incomplete Si-burning in nucleosynthesis terms, i.e. O (and C and lighter nuclei) are burned, Si and intermediate mass nuclei are the dominant products, and Fe-group elements can already be produced to some extent. It can be seen that there is a non-negligible amount of protons produced. These originate mostly from $(\alpha,p)$ reactions. Such reactions are not included in the reduced alpha-network, which only contains $(\alpha,\gamma)$ reactions for intermediate mass elements and beyond. Besides producing protons, these reactions also proceed faster than $(\alpha,\gamma)$ reactions, permitting a faster build-up of Fe-group nuclei. Additional reaction channels also permit a more complete burning of oxygen. Therefore, the amounts of (mostly burned) oxygen are overpredicted (upper panel), the dominant
intermediate mass elements show rather small differences (middle panel), while $^{52}$Fe and $^{56}$Ni are significantly underproduced (bottom panel). This particle was selected because it gives an extreme difference in $^{56}$Ni; particles which experience higher temperatures and complete Si-burning will produce significant and comparable amounts of Ni in the reduced alpha as well as the postprocessing network. Thus, the amount of $^{56}$Ni obtained in the simulations can be used as a reliable lower estimate.

\subsection{Gravitational radiation from the $0.7\msun +0.7\msun$~and 
$1.06\msun +0.81\msun$~white dwarfs collision}
\label{gwave}
Under the light of a recent work \citep{katz2012}, the head-on collision of WDs might not be as rare as is generally believed. If that is the case, the rate of WDs collisions may have an important impact in the sample of SNIa light-curves used in cosmology. Therefore, there is a necessity of clearly identify outliers from the sample. In order to do that, a gravitational wave detection will be a clear distinct signature that will differentiate among head-on collisions and other channels of supernova production \citep{loren2010}. With that in mind we calculated the gravitational emission of cases 2D7755 and 2D10855, using the reduced quadrupole approximation. Axisymetric scenarios have only one polarization in the gravitational wave emission \citep{finn1990}. In order to evaluate gravitational radiation in 2D-cylindrical SPH we need to calculate the reduced quadrupole in cylindrical coordinates and  integrate over the azimuthal angle \citep{cabezon2010}. For example, for the $xx$ component we have,

\begin{align}
\rquad_{xx} &=\int_V\rho\left(xx-\frac{1}{3}\mathbf{R}^2\right)dV\nonumber\\
&=\int_V\rho\left[\left(r\cos \varphi \right)^2-\frac{1}{3}\left(r^2+z^2\right)\right]rdrdzd\varphi\nonumber\\
&=\frac{1}{6}\int_r\int_z2\pi r\rho \left(r^2-2z^2\right)drdz\,.
\label{ixxcyl}
\end{align}

As we can see in \cite{garciasenz2009}, $2\pi r\rho$ is the 2D density $\eta$, and multiplied by $drdz$ gives the mass element, which within the SPH technique is the mass of each particle. Hence, we can write a discretized version of Eq.~\ref{ixxcyl} as

\begin{equation}
\rquad_{xx}=\frac{1}{6}\sum_{i=1}^Nm_i\left(r^2_i-2z^2_i\right)\,,
\label{ixxsph}
\end{equation}

\noindent
where the summation is extended over all the particles of the system. Following the same procedure we can find the following relations,

\begin{align}
\rquad_{xx} &=\rquad_{yy}=-\frac{1}{2}\rquad_{zz}\,,\\
\rquad_{xy} &=\rquad_{xz}=\rquad_{yz}=0\,.
\end{align}

Therefore, for an observer located along the Y axis ($\theta=\phi=\frac{\pi}{2}$) at a distance $D=10$~kpc we obtain the gravitational wave amplitudes for both polarizations,

\begin{align}
h_+ &=\frac{-3}{D}\frac{G}{c^4}\rquaddot_{xx}\,,\\
h_\times&=0\,.
\end{align}

There are different direct methods for evaluating the components of $\rquaddot_{xx}$. Nevertheless, time derivatives cause numerical difficulties due to two main reasons: the numerical noise introduced by discretization and the magnification of the high-frequency components of the noise. To avoid these problems \cite{finn1990} proposed three methods which avoid direct time derivatives of $\rquad$ to evaluate $\rquadot$ and $\rquaddot$: the ``momentum divergence formula" that uses the continuity equation to re-express the first time derivative of $\rquad$ with no explicit time derivatives; the ``first moment of momentum density" expression for $\rquadot$ that weights the mass contribution by the moment arm $r$ (rather than $r^2$); and the ``stress formula" that uses the stress tensor to derive an expression of $h_{lm}^{TT}$ that involves accelerations and no explicit time derivatives. In this work we use the method proposed by \cite{centrella1993} which is similar to the ``stress formula" of Finn and Evans, and at the same time takes advantage of the Lagrangian nature of SPH.

\begin{equation}
\rquaddot_{xx}=\frac{1}{3}\sum_{i=1}^Nm_i\left[\dot{r}^2_i+r_i\ddot{r}_i-2\left(\dot{z}^2_i+z_i\ddot{z}_i\right)\right]\,.
\end{equation}

The results show a single strong burst of gravitational emission, followed by a slow fade out of the signal (see Fig.~\ref{fig10}). The peak value is rather high for events happening in our galaxy, even comparable to those obtained by neutron star mergers at the Mpc scale, but with a much longer timescale. Nevertheless, being a single event laking of periodicity, its detection will be difficult. The long-term relaxation of the emission can span over several hundreds (even a thousand) of seconds, but without any trace of a ring-down phase, as expected from a system where there is no remnant. Figure~\ref{fig10} shows also the gravitational emission of the 2D10855 case. The emission in this scenario is considerably stronger, in spite of showing very similar properties compared to that of scenario 2D7755. The lack of symmetry in the emission of this case is due, as expected, to the different masses of the two WDs involved in this collision.

\section{Conclusions}
\label{conclusions}

In this paper we have studied the head-on collision scenario for white dwarfs. These events might represent a possible channel to type Ia supernovae because they trigger energetic explosions and yield reasonable amounts of $^{56}$Ni, compatible in most cases with faint supernova events. We presented seven simulations that explore mainly two parameters: initial mass of the WDs, and initial nuclear composition, this last one never being studied before.  We conducted a 2D versus 3D comparison and a 2D versus 2D with twice the number of particles, as convergence tests for both spatial and mass resolutions. In both cases we obtained a very good agreement, especially in the 2D comparisons, where the mass resolution was tested and showed a convergence for both, nucleosynthetic and energetic output. We confirm the results of previous works, where the yields of $^{56}$Ni increase with the initial mass of the system, obtaining $0.25\msun$ of $^{56}$Ni for the less massive scenario, and $1.02\msun$ for the super-Chandrasekhar simulation. The final kinetic energy of the fluid, in all cases is well beyond $10^{51}$~erg, making these collisions an interesting alternative to the standard scenarios to reproduce SNIa explosions. 

The dependence of the outcome with the initial nuclear composition is more complicated. In this case, there are two mechanisms acting in opposite directions. When the initial carbon abundance is low, the detonation happens at higher densities and temperatures, triggering high yields of $^{56}$Ni. However, having less $^{12}$C produces a smaller amount of IME and leads to a partial suppression of the yields. This combination between strength of the detonation and initial fuel, leads to a non-linear dependence of the final yields with the initial nuclear composition. 

In Table~\ref{table4} there is a comparison of the yields of $^{56}$Ni obtained in this work with those of other groups. We summarize there the main available information about the head-on collision of white dwarfs with masses as close as possible to those considered in these simulations.

Concerning calculations carried out with the SPH technique, for the collision of two $0.6\msun$ white dwarfs, \cite{rosswog2009.2} obtained around $0.32\msun$ of $^{56}$Ni, after post-processing the hydrodynamic output. A similar amount of $^{56}$Ni was reported in \cite{raskin2009} using an $\alpha$-network of 13 nuclei. A larger amount of radioactive nickel, $0.51\msun$ was, however, found in \cite{raskin2010}, this time with a pair of $0.64\msun$ white dwarfs (Table~\ref{table4}). 

A calculation carried out with a different hydrodynamic code has been reported by \cite{rosswog2009.2}. Using the FLASH AMR code, the amount of radioactive nickel ejected after the collision of two $0.6\msun$ WDs was considerably lower ($0.16\msun$) than the yields obtained with the SPH method. Nevertheless, \cite{hawley2012} have also recently studied the head-on scenario with FLASH, and they reported higher yields of $^{56}$Ni ($0.32\msun$, for $0.64+0.64$ WDs), more compatible with our results, stating that the discrepancy with the previous work might be due to time-step handling during the combustion stage. 

Three points may explain these differences: 1) resolution issues, 2) the treatment of the nuclear reactions, and 3) a possible dependence on the numerical technique used for calculating the hydrodynamical evolution. 

Resolution issues, can be related either with spatial, temporal and/or mass resolution. A lack of spatial resolution can lead to a poor description of the fluid properties in sensitive regions, like the shock front. A poor temporal resolution makes more difficult the integration of the equations for the nuclear reactions, losing precision or obtaining even non-physical results. And low mass resolution produces very massive fluid elements that, when undergoes nuclear burning, always assume homogeneous burning and generate an enormous amount of energy that can artificially trigger a too-early detonation. In this work we used a 2D axisymmetric SPH code ({\sl AxisSPH}) that provides a spatial resolution 5 times better than the standard 3D calculations of this scenario done so far. The price to pay is that we only can simulate the $b=0$ collision (i.e. pure head-on), which, based on previous studies, has to be taken as a limiting case, where the energetics and $^{56}$Ni production are maximal. This imposed symmetry, allowed us to decrease considerably the number of particles involved in the calculation without losing spatial resolution. In this way, we could put a stronger computational effort in the coupled nuclear network that evaluates the abundances and energetic evolution of the fluid due to nuclear reactions. 

The nuclear network and its implementation is the second suspect. In our case, we used an implicit 14-nuclei alpha-network that solves the equations of the nuclear species evolution and the temperature change, consistently. We used the operator splitting technique to allow the network to reach the small time steps it needs, which are computationally prohibitive for the hydrodynamic step. In this way, we can calculate the whole evolution with high temporal resolution, from the normal burning to the NSE, and the posterior freeze-out. In order to check the outcome of our reduced nuclear network, we have post-processed over 1000 particles for model 2D7755 and obtained 0.40~$M_\odot$ of nickel compared to the 0.36~$M_\odot$ resulting from the alpha network in the hydrodynamical simulations. This difference is due to the many other channels to burn the material up to nickel in the extended network, that are absent in the reduced one. Among these, for example, $(\alpha,p)$ reactions rates are significantly faster than the $(\alpha,\gamma)$ reactions for elements heavier than Ca. Therefore, the amount of nickel provided by the simulations may be underestimated by a 10\% and can be used as a lower limit.

The last possibility is that there is a dependence with the hydrodynamic code employed to do the calculation. There is an inherent difficulty for hydrocodes to trigger a detonation, because, nowadays, it is not possible to resolve the microinstabilities which might be the real onset of the detonation. Therefore, the promptness with which a hydrodynamic code can detonate relays on how efficiently it can reach sustained detonation conditions within the employed resolution. For example, as stated in \cite{raskin2010}, the conditions to have a sustained detonation in SPH are more difficult to fulfill than in mesh codes. In order to trigger the detonation a particle that reaches the proper conditions has to be capable of delivering enough energy to its neighbors. This leads to a delay in the detonation, which then occurs at more extreme conditions, enhancing the nuclear output. Nevertheless, the results of \cite{hawley2012}, that used FLASH to perform their simulations, are very similar to those obtained in this work, even to the 3D7755 low resolution calculation, pointing to the fact that the code dependence might be lower than previously thought, and that the real sources for the spread of the results are either resolution or the nuclear network.

In spite of the scatter in the specific value, all the calculations point to the fact that significant amounts of $^{56}$Ni are produced naturally by WDs collisions. Overall, this scenario has interesting properties to explain both faint supernova explosions, but also very bright events. The rate at which these collisions may occur in nature is still under study, but recent developments makes them an exciting possibility that deserves further research.

\section*{Acknowledgements}

The authors acknowledge useful comments of Cody Raskin and Frank Timmes concerning nucleosynthetic yields in the collisional scenario. D.G. and A.R. were supported by the Spanish MEC grants AYA2010-15685, AYA2008-04211-C02-C01, and DURSI of the Generalitat de Catalunya. R.M.C. was supported by the Swiss Platform for High-Performance and High-Productivity Computing within the {\em supernova} project. F.K.T was supported by the Swiss National Science Foundation (SNSF) and the EuroGENESIS and CompStar progams.  A.A. was supported by a Feodor Lynen Fellowship (Humboldt Foundation) combined with support from SNSF, and with the Helmholtz-University Young Investigator grant No. VH-NG-825.

\bibliographystyle{mn2e}
\bibliography{bibliography_r}

\clearpage

\begin{figure*}
\includegraphics[width=\textwidth]{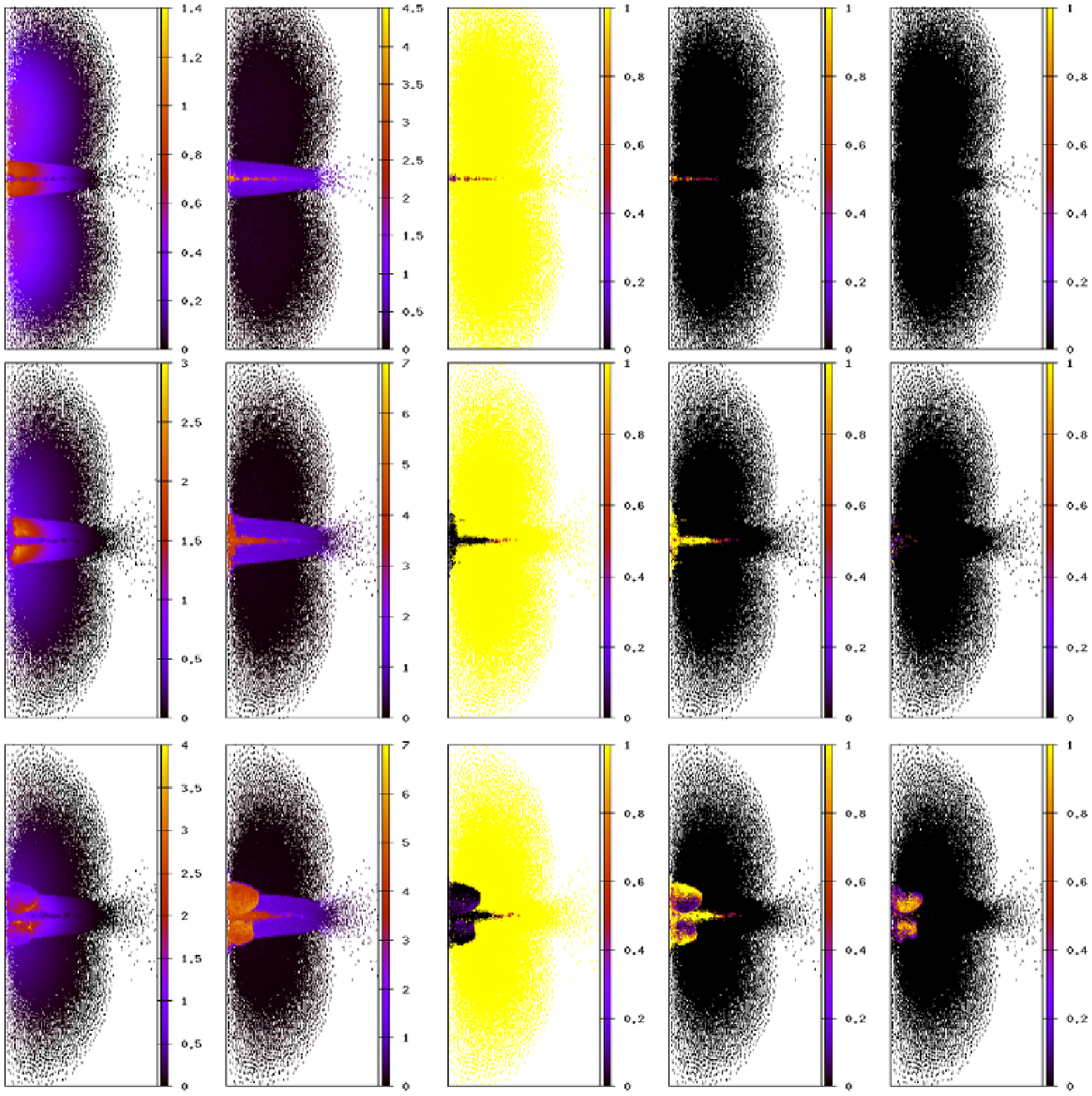}
\caption{Snapshots of the evolution of model 2D7755. From left to right, color scale indicates: density ($10^7$ g/cm$^3$), temperature ($10^9$ K), mass fraction of $^{4}$He $+^{12}$C $+^{16}$O, mass fraction of intermediate-mass elements (from $^{20}$Ne to $^{40}$Ca) and mass fraction of Fe-group elements (from $^{44}$Ti to $^{60}$Zn). From top to bottom: $t=5.63, 6.19, 6.30$~s. All boxes have the same size, $10^9$~cm in the X-direction and $2\cdot 10^9$~cm in the Y-direction.}
\label{fig02}
\end{figure*}
\clearpage

\begin{figure*}
\includegraphics[width=\textwidth]{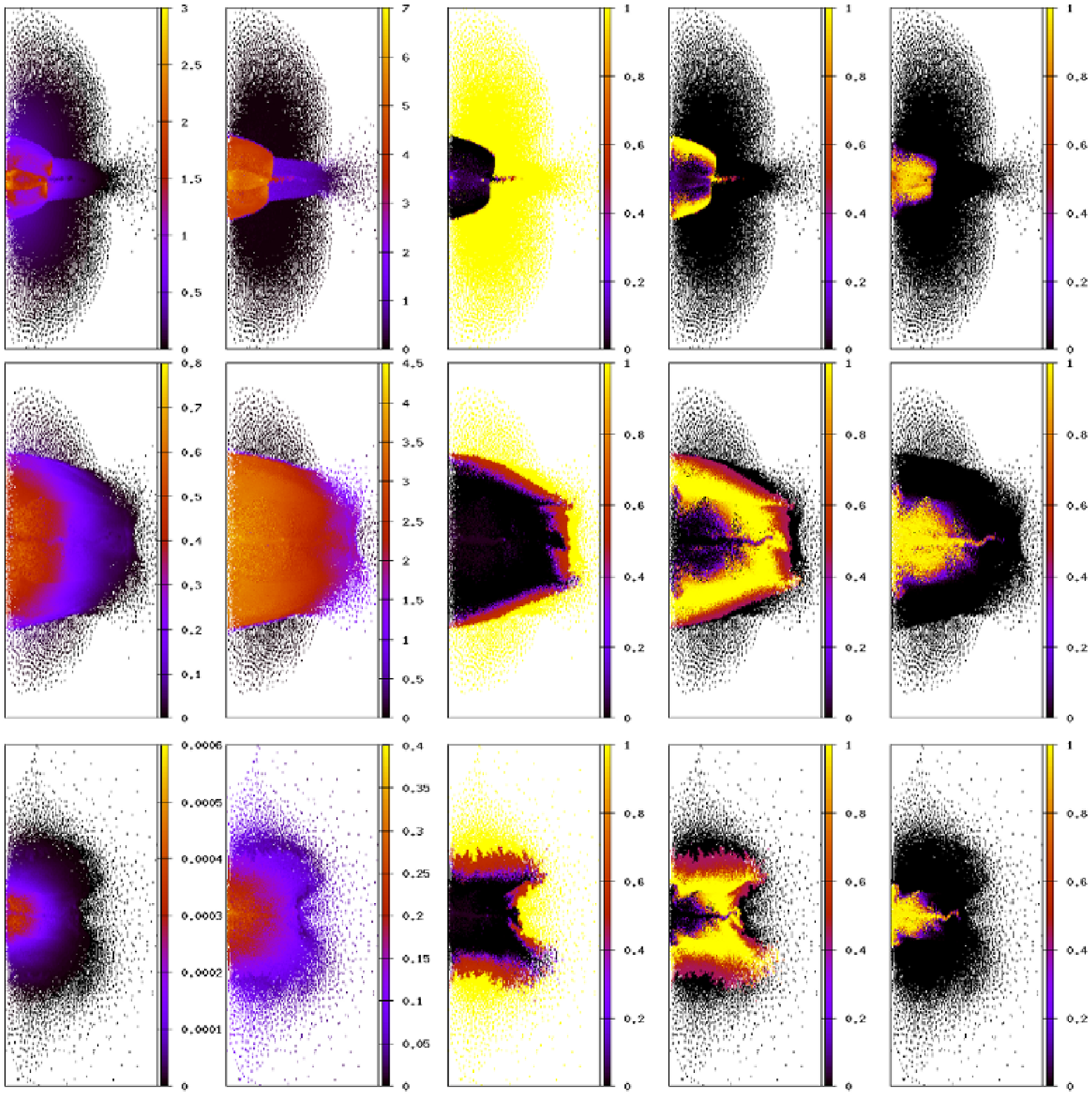}
\caption{Continuation of Fig.~\ref{fig02}. From top to bottom: $t=6.39, 6.80, 12.95$~s. Upper and middle row boxes size is $10^9$~cm in the X-direction and $2\cdot 10^9$~cm in the Y-direction. Lower row boxes are 20 times larger in each direction.}
\label{fig03}
\end{figure*}
\clearpage

\begin{figure}
\includegraphics[width=\columnwidth]{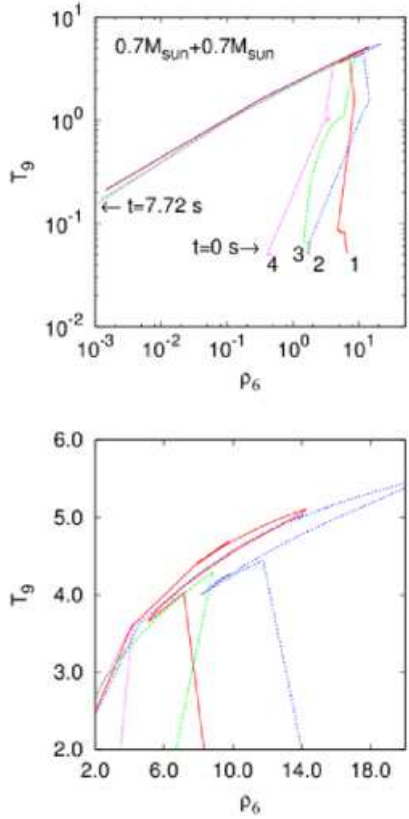}
\caption{Path followed by four selected particles in the $\rho-T$ plane for the $0.7\msun -0.7\msun$ collision (model 2D7755). Particles 1, 2, 3 and 4 belong to the upper white dwarf shown in Figs.~\ref{fig02} and \ref{fig03}. At $t=0$~s they were located at (comoving) coordinates (0, 0), (0, -0.5R$_{WD}$), (0.5R$_{WD}$, 0) and (0.5$R_{WD}$, -0.5R$_{WD}$) respectively, where $R_{WD}$ is the radius of the white dwarf. Point 1 is just at the center of the white dwarf. The lower panel shows a zoomed section of the high temperature and density region, where particles are shocked several times by the detonation fronts.}
\label{fig04}
\end{figure}

\begin{figure*}
\includegraphics[angle=-90,width=\textwidth]{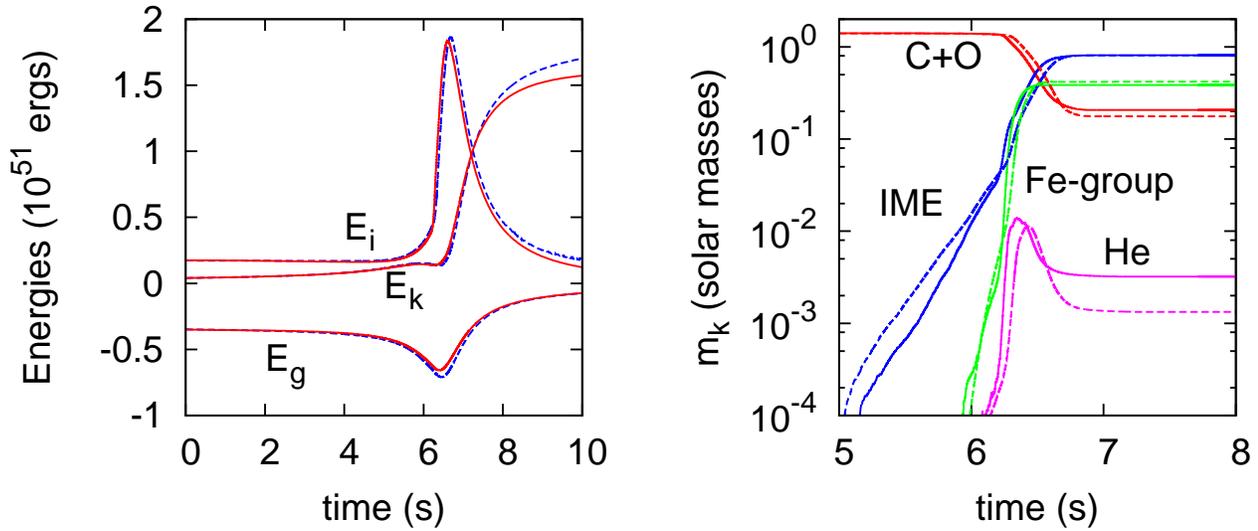}
\caption{Left, evolution of total internal energy ($\mathrm E_i$), kinetic energy ($\mathrm E_k$) and gravitational energy ($\mathrm E_g$) for models 2D7755 (solid lines)  and 3D7755 (dashed lines). Right, comparison of the abundances of four groups of elements:  $\alpha$ (pink), $^{12}$C$+^{16}$O (red),  IME (blue, from $^{20}$Ne~to $^{40}$Ca), and Fe-group (green, from $^{44}$Ti to $^{60}$Zn) for the {\sl AxisSPH}~(solid lines) and 3D calculations (dashed lines), respectively.}
\label{fig05}
\end{figure*}

\begin{figure*}
\includegraphics[width=\textwidth]{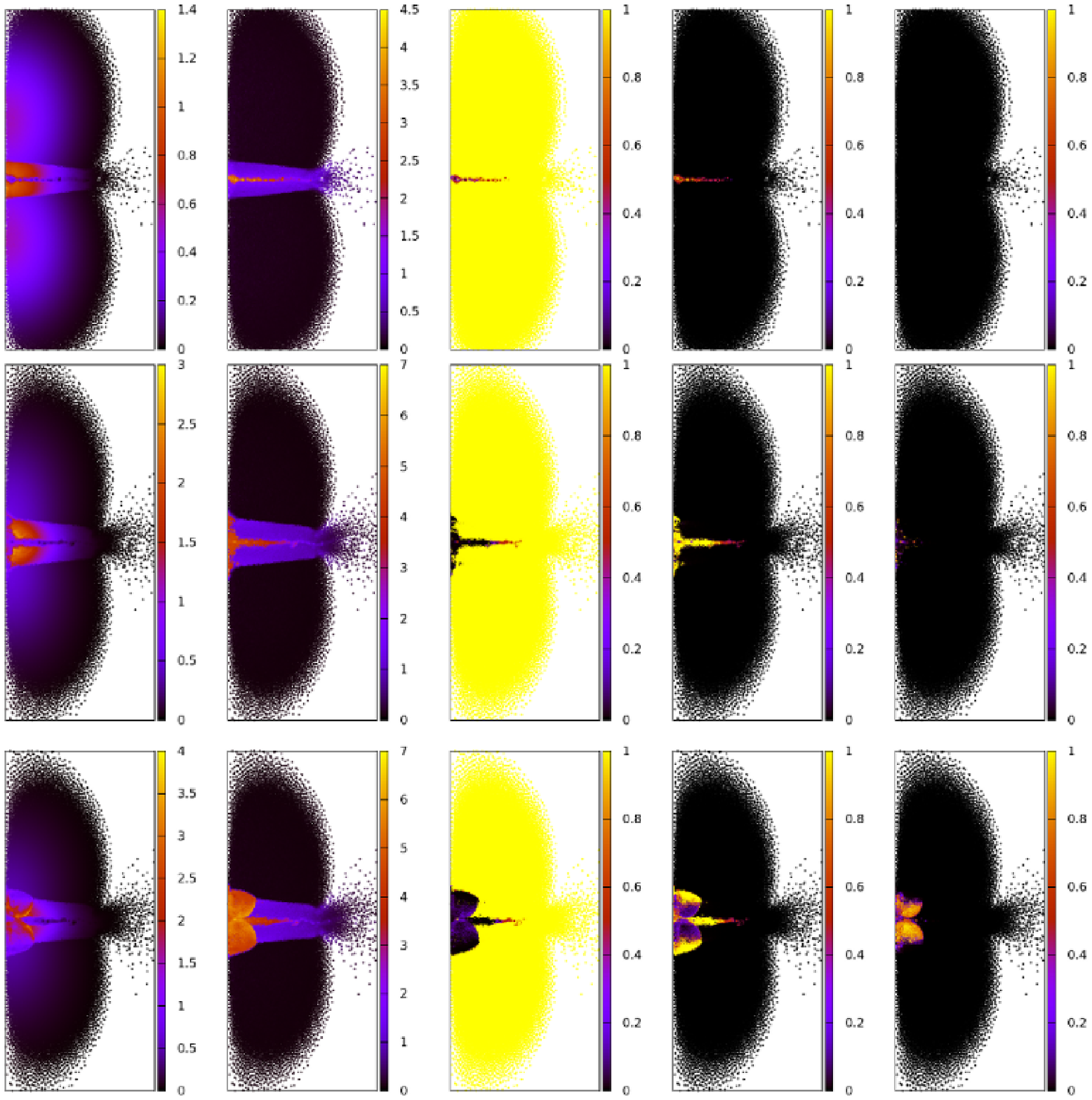}
\caption{Snapshots of the evolution of model 2D7755Res. From left to right, color scale indicates: density ($10^7$ g/cm$^3$), temperature ($10^9$ K), mass fraction of $^{4}$He $+^{12}$C $+^{16}$O, mass fraction of intermediate-mass elements (from $^{20}$Ne to $^{40}$Ca) and mass fraction of Fe-group elements (from $^{44}$Ti to $^{60}$Zn). From top to bottom: $t=5.64, 6.19, 6.30$~s. All boxes have the same size, $10^9$~cm in the X-direction and $2\cdot 10^9$~cm in the Y-direction.}
\label{fig02res}
\end{figure*}
\clearpage

\begin{figure*}
\includegraphics[width=\textwidth]{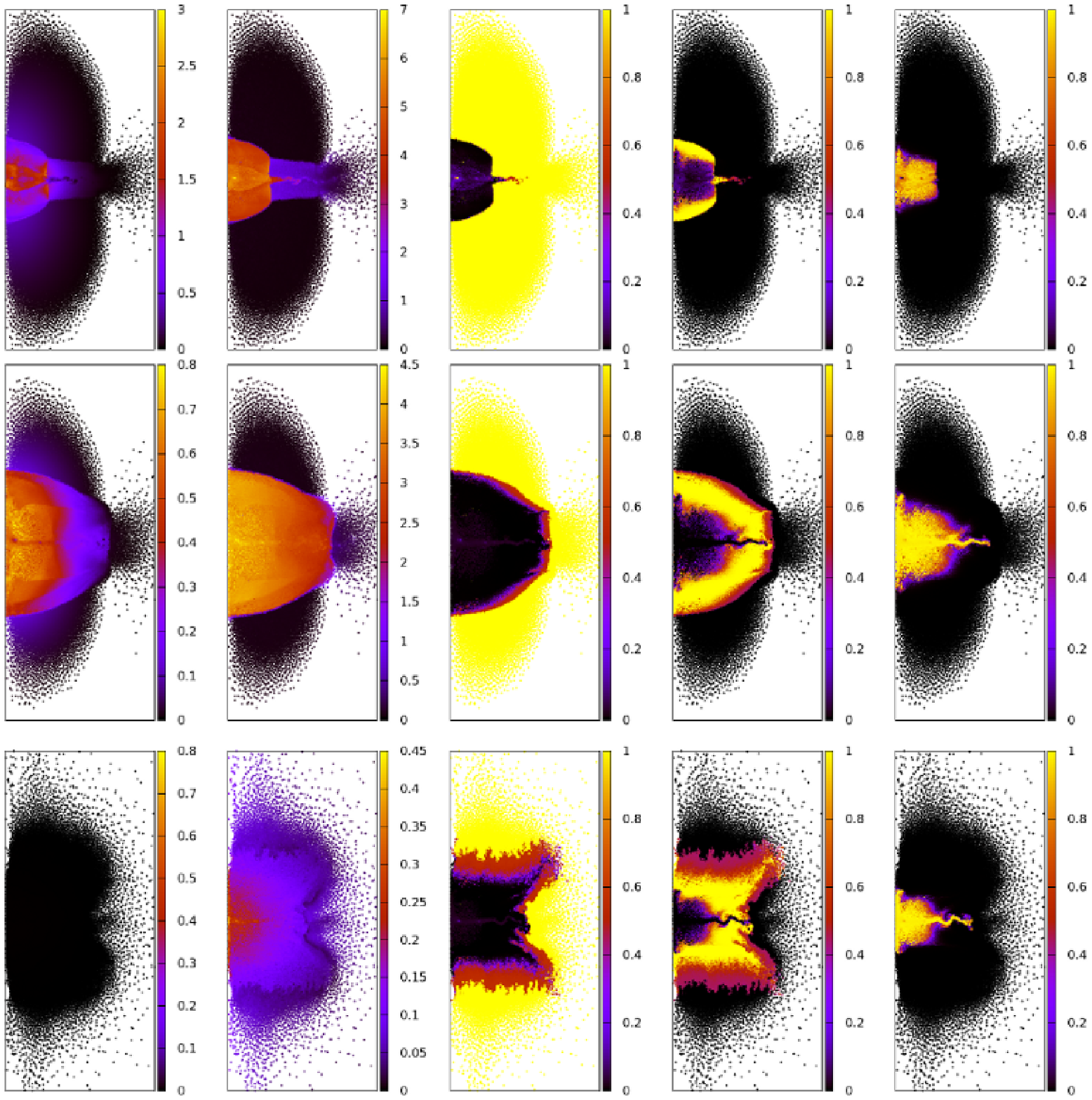}
\caption{Continuation of Fig.~\ref{fig02res}. From top to bottom: $t=6.39, 6.70, 13.42$~s. Upper and middle row boxes size is $10^9$~cm in the X-direction and $2\cdot 10^9$~cm in the Y-direction. Lower row boxes are 20 times larger in each direction.}
\label{fig03res}
\end{figure*}
\clearpage

\begin{figure*}
\includegraphics[width=\textwidth]{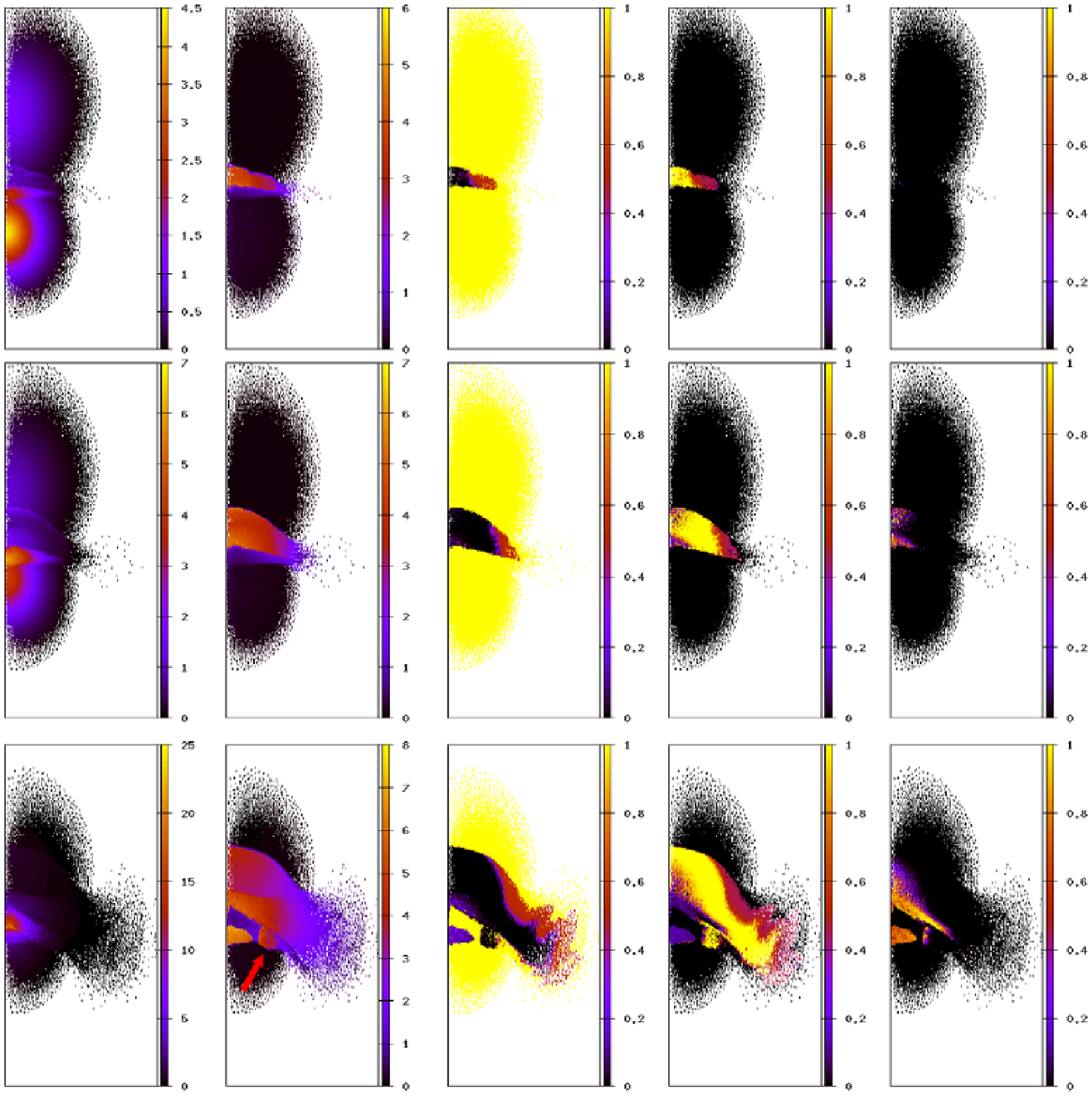}
\caption{Snapshots of the evolution of model 2D10855. From left to right, color scale indicates: density ($10^7$ g/cm$^3$), temperature ($10^9$~K), mass fraction of $^{4}$He $+^{12}$C $+^{16}$O, mass fraction of intermediate-mass elements (from $^{20}$Ne to $^{40}$Ca) and mass fraction of Fe-group elements (from $^{44}$Ti to $^{60}$Zn). From top to bottom: $t=16.87,17.12,17.53$~s. All boxes have the same size, $10^9$~cm in the X-direction and $2\cdot 10^9$~cm in the Y-direction. The red arrow in the last row points to the second detonation (see text for further details).}
\label{fig06}
\end{figure*}
\clearpage

\begin{figure*}
\includegraphics[width=\textwidth]{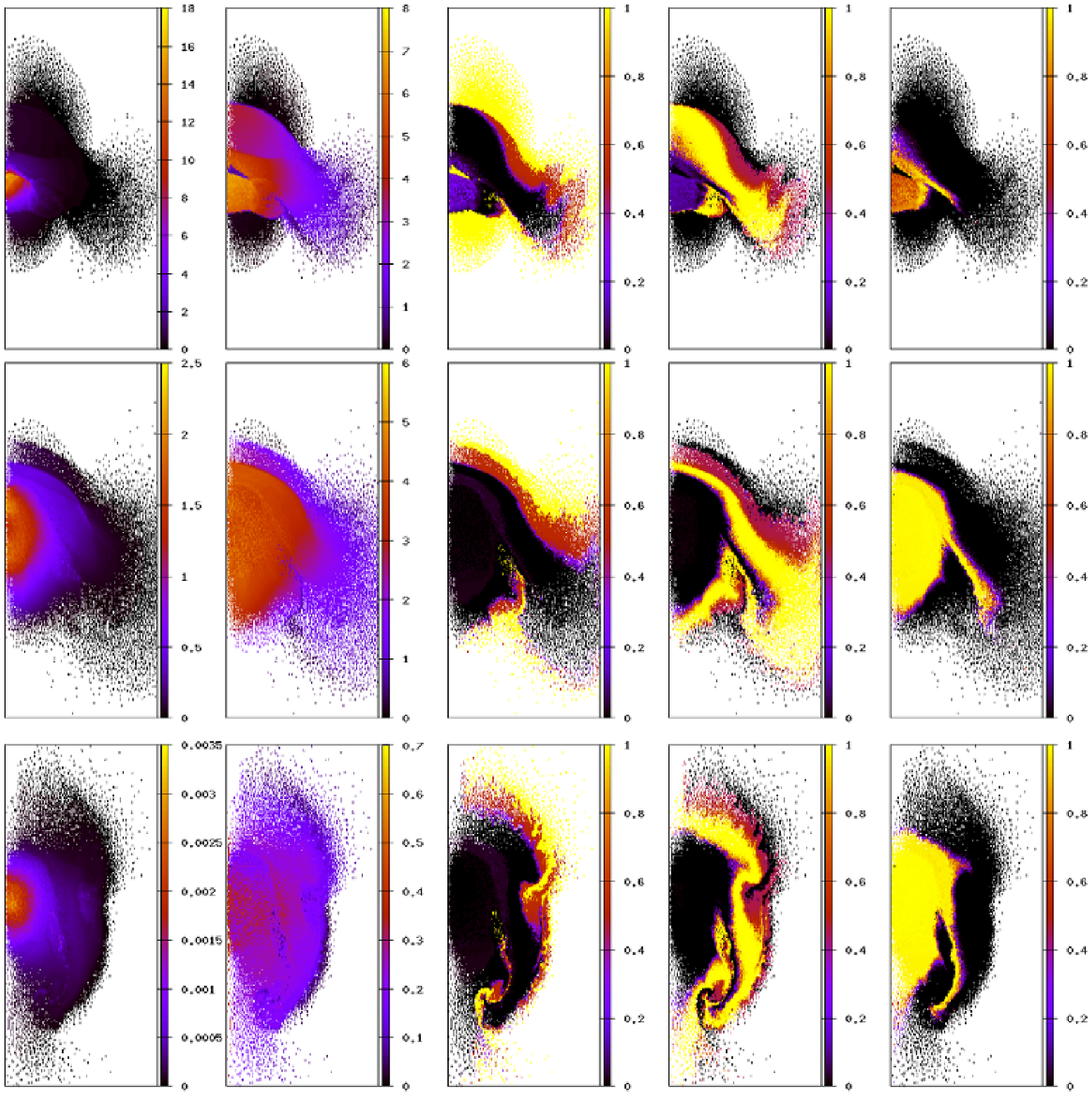}
\caption{Continuation of Fig.~\ref{fig06}. From top to bottom: $t=17.59,17.90,21.66$~s. Upper and middle row boxes size is $10^9$~cm in the X-direction and $2\cdot 10^9$~cm in the Y-direction. Lower row boxes are 10 times larger in each direction.}
\label{fig07}
\end{figure*}
\clearpage

\begin{figure*}
\includegraphics[angle=-90,width=\textwidth]{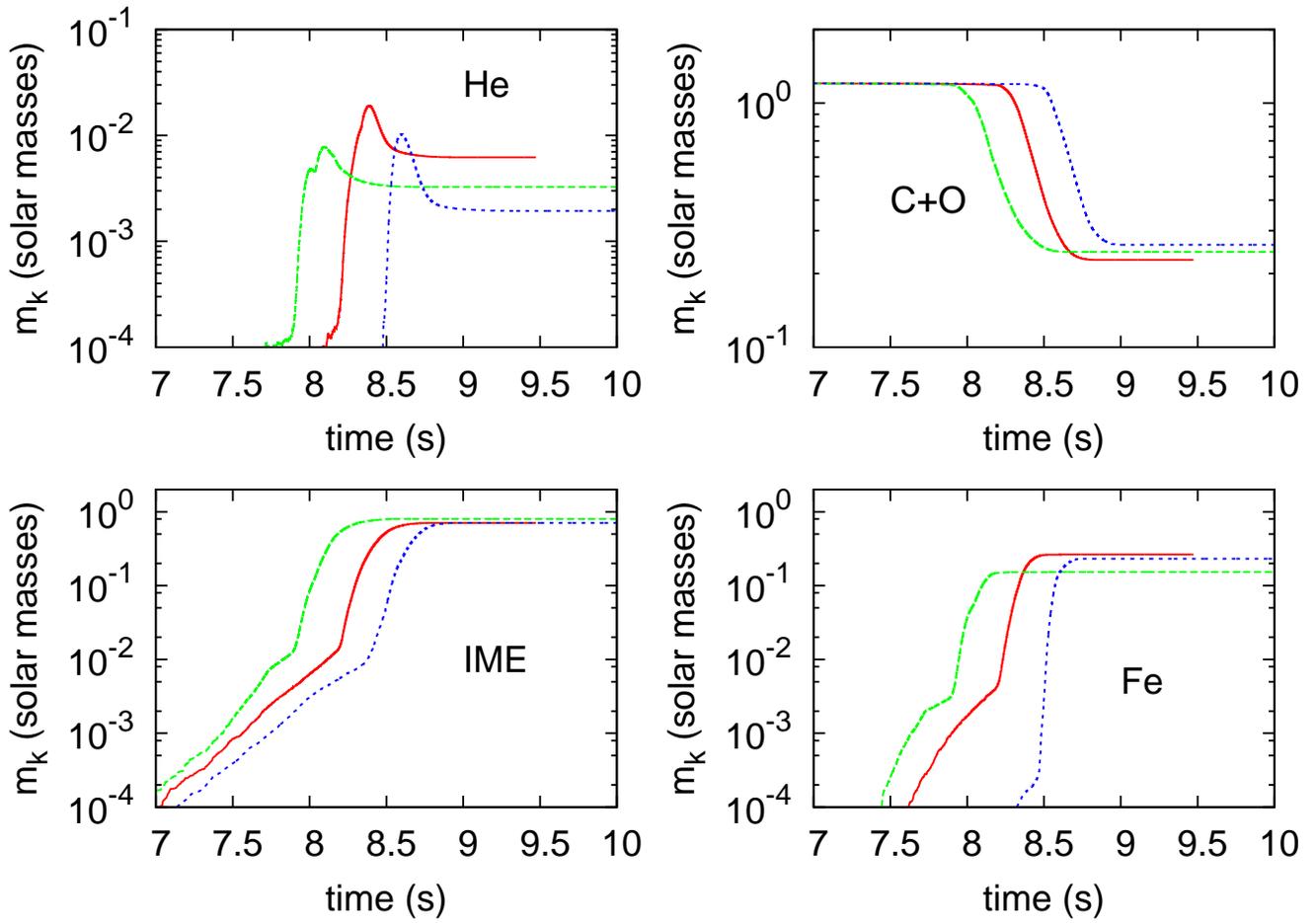}
\caption{Evolution of the abundances (of four groups of isotopes: $\alpha$, $^{12}$C+$^{16}$O, IME (from $^{20}$Ne to $^{40}$Ca) and Fe-group (from $^{44}$Ti to $^{60}$Zn)) calculated with {\sl AxisSPH}~as a function of the initial carbon abundance. Continuum lines are for the canonical case $\xc =0.5$, and dashed and dotted lines refer to $\xc=0.7$ and $\xc=0.3$, respectively.}
\label{fig08}
\end{figure*}
\clearpage

\begin{figure*}
\includegraphics[width=0.5\textwidth]{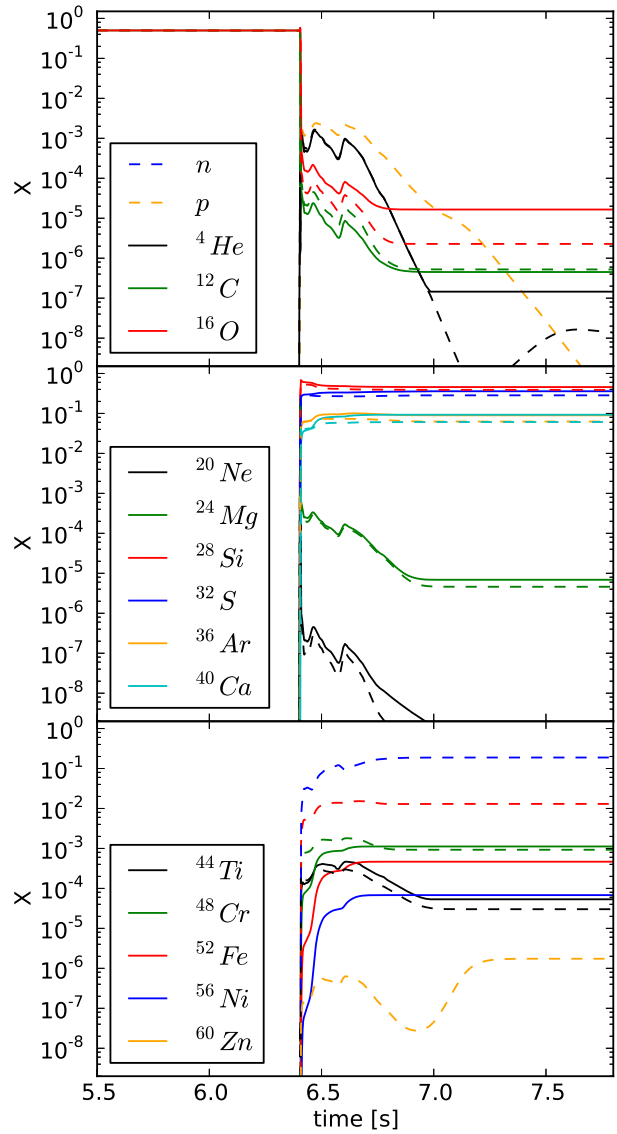}
\caption{Evolution of abundances for an example particle based on a reduced alpha network (solid lines) and on a extended network (dashed line). Different panels correspond to various groups of elements as indicated in the legends.}
\label{fig09} 
\end{figure*}
\clearpage

\begin{figure*}
\includegraphics[angle=-90,width=\columnwidth]{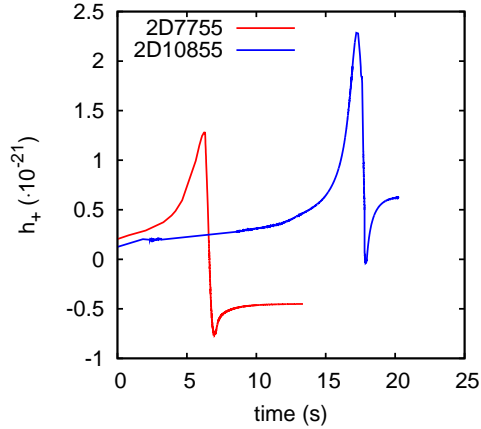}
\caption{h$_+$-polarization of the gravitational emission from models 2D7755 and 2D10855, as seen by an observer in the collision plane and at a distance $D=10$ kpc.}
\label{fig10}
\end{figure*}

\begin{table*}
\begin{center}
\begin{tabular}{|l|ccccccccc|}
\hline
\mc{1}{c}{Run} & $M_1~(\msunb)$ & $M_2~(\msunb)$&d$_{12}^0$ (km)& $v_1^0$ (km/s) & $v_2^0$ (km/s) & $\text{X}_{^{12}\text{C}} // \text{X}_{^{16}\text{O}}$ & Type & N& Res. (km)$^{(1)}$\\
\hline
\hline
\textbf{3D7755}     & 0.7  & 0.7  & 32,000 & -1,700 & 1,700 & 0.5 // 0.5 & 3D & 200,000 & 150 \\
\textbf{2D7755}     & 0.7  & 0.7  & 32,000 & -1,700 & 1,700 & 0.5 // 0.5 & 2D & 88,560  & 40 \\
\textbf{2D7755Res}  & 0.7  & 0.7  & 32,000 & -1,700 & 1,700 & 0.5 // 0.5 & 2D & 177,120 & 29 \\
\textbf{2D6655}     & 0.6  & 0.6  & 36,000 & -1,487 & 1,487 & 0.5 // 0.5 & 2D & 88,560  & 46 \\
\textbf{2D6673}     & 0.6  & 0.6  & 36,000 & -1,487 & 1,487 & 0.7 // 0.3 & 2D & 88,560  & 46 \\
\textbf{2D6637}     & 0.6  & 0.6  & 36,000 & -1,487 & 1,487 & 0.3 // 0.7 & 2D & 88,560  & 46 \\
\textbf{2D10855}    & 1.06 & 0.81 & 69,600 & -1,156 & 1,510 & 0.5 // 0.5 & 2D & 88,560  & 24 \\
\hline
\end{tabular}
\caption{Settings of the runs performed in this work. The name of the run states the type of simulation, the mass of the WDs and the initial nuclear composition in mass 
fractions of $^{12}$C and $^{16}$O. Comments: $^{(1)}$~this is the maximum resolution before the impact. During the collision there is a factor  $\simeq 2.5$~of improvement in the maximum resolution achieved in the 2D simulations.}  
\label{table1}
\end{center}
\end{table*}

\begin{table*}
\begin{center}
\begin{tabular}{|l|ccccc|}
\hline
\mc{1}{c}{Run} &KE $(\infty)$ $(10^{51}$~erg)&$\alpha~(\msunb)$&$^{12}$C+$^{16}$O~$(\msunb)$&IME~$(\msunb)$& Fe-group~$(\msunb)$\\
\hline
\hline
\textbf{3D7755}    & 1.71 & 0.0013 & 0.177 & 0.81 & 0.42 \\
\textbf{2D7755}    & 1.62 & 0.0032 & 0.205 & 0.81 & 0.39 \\
\textbf{2D7755Res} & 1.63 & 0.0026 & 0.197 & 0.82 & 0.39 \\
\textbf{2D6655}    & 1.33 & 0.0062 & 0.228 & 0.71 & 0.27 \\
\textbf{2D6673}    & 1.34 & 0.0033 & 0.245 & 0.80 & 0.15 \\
\textbf{2D6637}    & 1.10 & 0.0020 & 0.263 & 0.71 & 0.23 \\
\textbf{2D10855}   & 2.06 & 0.0081 & 0.195 & 0.63 & 1.05 \\
\hline
\end{tabular}
\caption{Kinetic energy at infinity and group abundances of the calculated models after the freezing of nuclear reactions.}
\label{table2}
\end{center}
\end{table*}

\begin{table*}
\begin{center}
\begin{tabular}{|c|c|c|c|c|c|c|c|}
\hline
\hline
~ & 3D7755~$(\msunb)$ & 2D7755~$(\msunb)$ & 2D7755Res~$(\msunb)$ & 2D6655~$(\msunb)$ & 2D6673~$(\msunb)$ & 2D6637~$(\msunb)$ & 2D10855~$(\msunb)$\\
\hline 
\hline

$\alpha$   &$1.33\cdot10^{-3}$ &$3.20\cdot10^{-3}$ &$2.58\cdot10^{-3}$ &$6.22\cdot10^{-3}$ &$3.27\cdot10^{-3}$ &$1.93\cdot10^{-3}$ &$8.14\cdot10^{-3}$\\
$^{12}$C   &$5.20\cdot10^{-2}$ &$5.50\cdot10^{-2}$ &$5.09\cdot10^{-2}$ &$6.60\cdot10^{-2}$ &$1.08\cdot10^{-1}$ &$3.91\cdot10^{-2}$ &$4.37\cdot10^{-2}$\\
$^{16}$O   &$1.25\cdot10^{-1}$ &$1.52\cdot10^{-1}$ &$1.46\cdot10^{-1}$ &$1.62\cdot10^{-1}$ &$1.37\cdot10^{-1}$ &$2.24\cdot10^{-1}$ &$1.49\cdot10^{-1}$\\
$^{20}$Ne  &$1.28\cdot10^{-4}$ &$3.20\cdot10^{-5}$ &$2.59\cdot10^{-5}$ &$9.88\cdot10^{-5}$ &$1.20\cdot10^{-4}$ &$5.90\cdot10^{-4}$ &$1.15\cdot10^{-3}$\\ 
$^{24}$Mg  &$2.45\cdot10^{-2}$ &$2.86\cdot10^{-2}$ &$2.70\cdot10^{-2}$ &$3.01\cdot10^{-2}$ &$5.85\cdot10^{-2}$ &$2.43\cdot10^{-2}$ &$4.34\cdot10^{-2}$\\
$^{28}$Si  &$3.76\cdot10^{-1}$ &$4.08\cdot10^{-1}$ &$4.10\cdot10^{-1}$ &$3.67\cdot10^{-1}$ &$4.33\cdot10^{-1}$ &$3.33\cdot10^{-1}$ &$3.21\cdot10^{-1}$\\
$^{32}$S   &$2.43\cdot10^{-1}$ &$2.30\cdot10^{-1}$ &$2.33\cdot10^{-1}$ &$1.96\cdot10^{-1}$ &$1.94\cdot10^{-1}$ &$2.17\cdot10^{-1}$ &$1.65\cdot10^{-1}$\\
$^{36}$Ar  &$6.94\cdot10^{-2}$ &$6.10\cdot10^{-2}$ &$6.24\cdot10^{-2}$ &$4.98\cdot10^{-2}$ &$5.00\cdot10^{-2}$ &$5.81\cdot10^{-2}$ &$4.16\cdot10^{-2}$\\
$^{40}$Ca  &$9.50\cdot10^{-2}$ &$8.20\cdot10^{-2}$ &$8.37\cdot10^{-2}$ &$6.29\cdot10^{-2}$ &$6.60\cdot10^{-2}$ &$7.39\cdot10^{-2}$ &$5.35\cdot10^{-2}$\\
$^{44}$Ti  &$4.85\cdot10^{-4}$ &$4.20\cdot10^{-4}$ &$3.58\cdot10^{-4}$ &$2.97\cdot10^{-4}$ &$1.65\cdot10^{-4}$ &$2.25\cdot10^{-4}$ &$2.20\cdot10^{-3}$\\
$^{48}$Cr  &$4.34\cdot10^{-3}$ &$3.80\cdot10^{-3}$ &$3.93\cdot10^{-3}$ &$2.85\cdot10^{-3}$ &$2.80\cdot10^{-3}$ &$3.02\cdot10^{-3}$ &$3.67\cdot10^{-3}$\\
$^{52}$Fe  &$2.26\cdot10^{-2}$ &$1.90\cdot10^{-2}$ &$2.10\cdot10^{-2}$ &$1.07\cdot10^{-2}$ &$9.29\cdot10^{-3}$ &$1.28\cdot10^{-2}$ &$2.45\cdot10^{-2}$\\
$^{56}$Ni  &$3.91\cdot10^{-1}$ &$3.62\cdot10^{-1}$ &$3.64\cdot10^{-1}$ &$2.49\cdot10^{-1}$ &$1.34\cdot10^{-1}$ &$2.15\cdot10^{-1}$ &$1.02$\\
$^{60}$Zn  &$3.48\cdot10^{-4}$ &$4.02\cdot10^{-4}$ &$3.45\cdot10^{-4}$ &$3.17\cdot10^{-4}$ &$1.08\cdot10^{-4}$ &$3.22\cdot10^{-4}$ &$1.64\cdot10^{-3}$\\
\hline
\end{tabular}
\caption{Detailed final abundance of the 14 nuclei for models of Table~\ref{table1}.}
\label{table3}
\end{center}
\end{table*}

\begin{table*}
\begin{center}
\begin{tabular}{|c|c|c|c|c|c|}
\hline
\hline
 Calculation & Masses~$(\msunb)$ & Code &Particles (Res., -km-)$^{(1)}$&  Nuclear Network & $^{56}$Ni~$(\msunb)$ \\
\hline 
\hline
\cite{benz1989}                       & 0.60+0.60  & 3D-SPH & $5,000$~($-$)       & 14-isotope$^{(2)}$                                                   & 0.008 \\
\hline
\multirow{2}{*}{\cite{rosswog2009.2}} & 0.60+0.60  & FLASH  & $-$ (49)              & 19-isotope$^{(3)}$                                                   & 0.16 \\
                                      & 0.60+0.60  & 3D-SPH & $2\cdot 10^6$~($-$) & 7-isotope$^{(4)}$ + post-processing$^{(3)}$                          & 0.32 \\
\hline
\cite{raskin2009}                     & 0.60+0.60  & 3D-SPH & $8\cdot 10^5$~($-$) & 13-isotope$^{(3)}$                                                   & 0.34 \\
\hline
\multirow{3}{*}{\cite{raskin2010}}    & 0.64+0.64  & 3D-SPH & $2\cdot 10^5$~($-$) & \multirow{3}{*}{13-isotope$^{(3)}$ + hybrid burning$^{(5)}$}         & 0.51 \\
                                      & 0.64+0.64  & 3D-SPH & $2\cdot 10^6$~($-$) &                                                                      & 0.53 \\
                                      & 1.06+0.81  & 3D-SPH & $2\cdot 10^6$~($-$) &                                                                      & 0.90 \\
\hline
\cite{hawley2012}                     & 0.64+0.64  & FLASH  & $-$  (130)              & 13-isotope$^{(3)}$                                                & 0.32 \\
\hline
\cite{kusnhir2013}                    & 0.60+0.60  & FLASH  & $-$  (8.5)             & \multirow{3}{*}{Not specified}                                       & 0.29 \\
                                      & 0.70+0.70  & FLASH  & $-$  (7.6)             &                                                                      & 0.50 \\
                                      & 1.00+0.80  & FLASH  & $-$  (6.8)             &                                                                      & 0.89 \\

\hline
\multirow{6}{*}{This work (2013)}     & 0.60+0.60  & 2D-SPH & $88,560$~(46)      & \multirow{4}{*}{14-isotope$^{(6)}$ + temperature coupling$^{(6)}$}   & 0.25 \\     
                                      & 0.70+0.70  & 3D-SPH & $2\cdot 10^5$~(150) &                                                                      & 0.39 \\     
                                      & 0.70+0.70  & 2D-SPH & $88,560$~(40)      &                                                                      & 0.36 \\
                                      & 0.70+0.70  & 2D-SPH & $177,120$~(29)     &                                                                      & 0.36 \\
                                      & 1.06+0.81  & 2D-SPH & $88,560$~(24)      &                                                                      & 1.02 \\
                                      &            &        &               &                                                                      &      \\
                                      & 0.70+0.70  & 2D-SPH & $88,560$~($-$)      & Post-processing$^{(7)}$                                              & 0.40 \\
\hline
\end{tabular}
\caption{Comparison of the outcomes for the $^{56}$Ni yields of a head-on white collision from previous works with those of this work. Comments: $^{(1)}$ \protect  for 
 grid-based codes the maximum resolution in km is shown as well as the maximum resolution at t=0 s in  SPH calculations when available; $^{(2)}$ \protect \cite{benz1989}; $^{(3)}$ \protect \cite{timmes1999}; $^{(4)}$ \protect \cite{hix1998}; $^{(5)}$ \protect \cite{raskin2010}; $^{(6)}$ \protect \cite{cabezon2004}; $^{(7)}$~\protect \cite{thielemann2011}}
\label{table4}
\end{center}
\end{table*}

\clearpage

\appendix

\section[]{Main features of AxisSPH}
\label{appAxisSPH}

The hydrocode {\sl AxisSPH} is a two-dimensional SPH code written in axisymmetric geometry which incorporates several improvements over existing versions of these hydrodynamic codes \citep{garciasenz2009}. Its main advantage relies in the more accurate treatment of particles moving close to the symmetry z-axis which is a singular line.  
 
All axisymmetric SPH codes assume that fluid mass elements are torus characterized by its mass $m$ and distance $r$ to the symmetry axis. The basic hydrodynamic SPH equations can be obtained from the 3D-SPH equations after a simple substitution of the interpolating kernel $W_{3D}$ by its two-dimensional counterpart $W_{2D}$ and using the following relationship between the volumetric $\rho$ and surface $\eta$ densities,  

\begin{equation}
\rho=\frac{\eta}{2\pi r}
\end{equation}

To avoid the underestimation of $\eta$ in the neighborhood of the symmetry axis it is usual to add reflective particles to the numerical scheme. Still, this is not sufficient to ensure the correct behavior of the dynamics near the z-axis because any error in the SPH interpolations may lead to a large error in $\rho$ when $r\rightarrow 0$. To elude that problem {\sl AxisSPH} makes use of a correction factor $f_1$ which reduces the value of $\eta$ when $\zeta=\frac{r}{h}\le 2$, enforcing $\eta\rightarrow 0$ when $r\rightarrow 0$. A similar function $f_2$ works to ensure the adequate behavior of the $r$-component of velocity near the symmetry axis. The ensuing basic Euler equations are (see \citealt{relano2012} for the detailed derivation of these equations):

\vskip 0.1 cm
\noindent {\sl Mass equation},

\begin{equation}
\label{surfacedensity}
\hat\eta_i=\sum_{j=1}^N m_j W_{ij}\cdot f_1^i=\eta_i\cdot f_1^i
\end{equation}

\noindent {\sl Momentum equations},

\begin{align}
\ddot r_i= & 2\pi\frac{P_i}{\hat\eta_i}\nonumber\\
 &-2\pi \sum_{j=1}^N \left[ m_j \left( \frac{P_i r_i}{\hat\eta_i^2} \cdot f_1^i(\zeta_i)+ \frac{P_j r_j}{\hat \eta _j^2}+ \Pi _{ij}^{2D} \right) \frac{\partial W_{ij}}{\partial r_i} \right]\nonumber \\
 &-\left(\frac{2\pi P_i r_i}{\hat\eta_i\cdot f_1^i}\right)\frac{df_1^i (\zeta_i)}{dr_i}\,,
\label{accel_r}
\end{align}

\begin{equation}
\label{accel_z}
\ddot z_i=-2\pi\sum_{j=1}^N m_j\left(\frac{P_i r_i}{\eta_i^2}\cdot f_1^i(\zeta_i)+\frac{P_j r_j}{\hat\eta_j^2}+\Pi_{ij}^{2D}\right)\frac{\partial W_{ij}}{\partial z_i}\,.
\end{equation}

\noindent {\sl Energy equation}, 

\begin{equation}
\label{energy}
\frac{du_i}{dt}=-2\pi\frac{P_i}{\hat\eta_i} v_{r_i}+2\pi\frac{P_i r_i}{\hat\eta_i^2}\frac{d\hat\eta_i}{dt}+\frac{1}{2}\sum_{j=1}^N m_j\Pi_{ij}^{2D} ({\bf v_i-v_j})\cdot {\bf\nabla_i^{2D}}W_{ij}\,,
\end{equation}

\noindent being

\begin{align}
\frac{d\hat\eta_i}{dt}=&\sum_{j=1}^N m_j (f_1^i v_{r_i}-f_2^i v_{r_j}) \frac{\partial W_{ij}}{\partial r_i}\nonumber\\
&+\sum_{j=1}^N m_j\left(\frac{\partial f_1^i}{\partial r_i} v_{r_i}-\frac{\partial f_2^i}{\partial r_i} v_{r_j}\right)W_{ij}\nonumber\\
&+f_1^i\sum_{j=1}^N m_j (v_{z_i}-v_{z_j})\frac{\partial W_{ij}}{\partial z_i}\,.
\label{continuity}
\end{align}

Where $\hat\eta$ is the corrected value of the surface density, $\Pi_{ij}^{2D}$ is the artificial viscosity and the remaining symbols have their usual meaning in SPH, \citep{monaghan1992}. The first terms on the right side of equations \ref{accel_r} and \ref{energy}, called the hoop-stress terms, are specific to the axisymmetric geometry and they can be a source of troubles if the surface density $\eta$ is not well evaluated. The correction factors, only valid for the cubic-spline interpolating kernel,  affecting the density $\eta$ and $v_{r_i}$ are,  

\begin{equation}
f_1^i(\zeta)
\begin{cases}
\left[\frac{7}{15}\zeta_i^{-1}+\frac{2}{3}\zeta_i-\frac{1}{6}\zeta_i^3+\frac{1}{20}\zeta_i^4\right]^{-1} & 0<\zeta_i\leq 1\,,\\
\\
\left[\frac{8}{15}\zeta_i^{-1}-\frac{1}{3}+\frac{4}{3}\zeta_i-\frac{2}{3}\zeta_i^2\right.\\
\left.+\frac{1}{6}\zeta_i^3-\frac{1}{60}\zeta_i^4\right]^{-1}& 1<\zeta_i\leq 2\,,\\
\\
1 & \zeta_i>2\,,
\end{cases}
\label{correctionf_1}
\end{equation}

\begin{equation}
f_2^i(\zeta)
\begin{cases}
\left[\frac{14}{15}\zeta_i^{-1}+\frac{4}{9}\zeta_i-\frac{1}{15}\zeta_i^3+\frac{1}{60}\zeta_i^4\right]^{-1} & 0<\zeta_i\leq 1\,,\\
\\
\left[\frac{-1}{45}\zeta_i^{-2}+\frac{16}{15}\zeta_i^{-1}-\frac{1}{3}+\frac{8}{9}\zeta_i\right.\\
\left.-\frac{1}{3}\zeta_i^2+\frac{1}{15}\zeta_i^3-\frac{1}{180}\zeta_i^4\right]^{-1} & 1<\zeta_i\leq 2\,,\\
\\
1 & \zeta_i>2\,,
\end{cases}
\label{correctionf_2}
\end{equation}

\noindent with $\zeta_i=\frac{r_i}{h_i}$. 

The basic scheme described above was conveniently modified to describe the hydrodynamics of  the encounter of two white dwarfs. Instead of the energy equation \ref{energy} we used a similar expression based on the temperature equation, which is more appropriate to handle degenerate structures such as white dwarfs. The rate of released nuclear energy obtained with the alpha-network was added to the right of the temperature equation. When $T> 5\cdot 10^8$~K and $\rho > 2\cdot 10^6$~g.cm$^{-3}$~the implicit coupling between the molar fractions of the species and the temperature equation is turned on, as described in \citet{cabezon2004}. Finally  self-gravity was added to equations \ref{accel_r} and \ref{accel_z} to compute the acceleration of the particle. In the present form of {\sl AxisSPH} gravity is calculated by direct particle-particle interaction, taking into account the ring-like structure of the mass-particles. While this  procedure is exact and easily parallelizable, the computational burden scales as $\sim N^2$, limiting in practice the total number of particles used to carry out the simulations.

\end{document}